\documentclass[twocolumn,a4paper,nofootinbib]{revtex4}
\usepackage{amsfonts,amsmath,amssymb,mathrsfs,dsfont,yfonts,bbm}
\usepackage[shortlabels]{enumitem}
\usepackage[normalem]{ulem}
\usepackage{tikz}
\usepackage{xcolor}
\usepackage{physics}
\begin{document}

\newcommand{\ad}[1] {
  \textcolor{blue}{#1}}

\newcommand{\Supertwistor}{\Cset \mathrm{P}^{3|4}}
\newcommand{\Twistorspace}{\Cset \mathrm{P}^{3}}
\newcommand{\half}{\frac{1}{2}}
\newcommand{\diff}{\mathrm{d}}
\newcommand{\ra}{\rightarrow}
\newcommand{\Zset}{{\mathbb Z}}
\newcommand{\Cset}{{\,\,{{{^{_{\pmb{\mid}}}}\kern-.47em{\mathrm C}}}}}
\newcommand{\Rset}{{\mathrm{I}\!\mathrm{R}}}
\newcommand{\gra}{\alpha}
\newcommand{\grl}{\lambda}
\newcommand{\gre}{\epsilon}
\newcommand{\zb}{{\bar{z}}}
\newcommand{\mn}{{\mu\nu}}
\newcommand{\Acal}{{\mathcal A}}
\newcommand{\Rcal}{{\mathcal R}}
\newcommand{\Dcal}{{\mathcal D}}
\newcommand{\Mcal}{{\mathcal M}}
\newcommand{\Ncal}{{\mathcal N}}
\newcommand{\Kcal}{{\mathcal K}}
\newcommand{\Lcal}{{\mathcal L}}
\newcommand{\Scal}{{\mathcal S}}
\newcommand{\CW}{{\mathcal W}}
\newcommand{\Bcal}{\mathcal{B}}
\newcommand{\Ccal}{\mathcal{C}}
\newcommand{\Vcal}{\mathcal{V}}
\newcommand{\Ocal}{\mathcal{O}}
\newcommand{\Zcal}{\mathcal{Z}}
\newcommand{\Zb}{\overline{Z}}
\newcommand{\Urm}{{\mathrm U}}
\newcommand{\Srm}{{\mathrm S}}
\newcommand{\SO}{\mathrm{SO}}
\newcommand{\Sp}{\mathrm{Sp}}
\newcommand{\SU}{\mathrm{SU}}
\newcommand{\U}{\mathrm{U}}
\newcommand{\be}{\begin{equation}}
\newcommand{\ee}{\end{equation}}
\newcommand{\Comment}[1]{{}}
\newcommand{\tQ}{\tilde{Q}}
\newcommand{\tq}{{\tilde{q}}}
\newcommand{\trho}{\tilde{\rho}}
\newcommand{\tphi}{\tilde{\phi}}
\newcommand{\Qcal}{\mathcal{Q}}
\newcommand{\tmu}{\tilde{\mu}}
\newcommand{\dbar}{\bar{\partial}}
\newcommand{\p}{\partial}
\newcommand{\eg}{{\it e.g.\;}}
\newcommand{\ie}{{\it i.e.\;}}
\newcommand{\twistor}{\Cset \mathrm{P}^{3}}
\newcommand{\note}[2]{{\footnotesize [{\sc #1}}---{\footnotesize   #2]}}
\newcommand{\CL}{\mathcal{L}}
\newcommand{\CJ}{\mathcal{J}}
\newcommand{\CA}{\mathcal{A}}
\newcommand{\CH}{\mathcal{H}}
\newcommand{\CD}{\mathcal{D}}
\newcommand{\CE}{\mathcal{E}}
\newcommand{\CQ}{\mathcal{Q}}
\newcommand{\CB}{\mathcal{B}}
\newcommand{\CC}{\mathcal{C}}
\newcommand{\CO}{\mathcal{O}}
\newcommand{\CT}{\mathcal{T}}
\newcommand{\CI}{\mathcal{I}}
\newcommand{\CN}{\mathcal{N}}
\newcommand{\CS}{\mathcal{S}}
\newcommand{\CM}{\mathcal{M}}

\newcommand{\bea}{\begin{eqnarray}}
\newcommand{\eea}{\end{eqnarray}}
\newcommand{\DZ}{\mathds{Z}}

\parskip 11pt
\title{\Large{Quantum Codes, CFTs, and Defects}}
\author{Matthew Buican$^{\dagger}$, Anatoly Dymarsky$^{*,\#}$, and Rajath Radhakrishnan$^{\dagger}$}
\affiliation{$^{\dagger}$CTP and Department of Physics and Astronomy \\ Queen Mary University of London, London E1 4NS, UK\\ $^{*}$Department of Physics and Astronomy, University of Kentucky\\ 505 Rose Street, Lexington, KY, 40506, USA\\ \# Skolkovo Institute of Science and Technology \\ Skolkovo Innovation Center, Moscow, Russia\\}

\begin{abstract}
\noindent
We give a general construction relating Narain rational conformal field theories (RCFTs) and associated 3d Chern-Simons (CS) theories to quantum stabilizer codes. Starting from an abelian CS theory with a fusion group consisting of $n$ even-order factors, we map a boundary RCFT to an $n$-qubit quantum code. When the relevant 't Hooft anomalies vanish, we can orbifold our RCFTs and describe this gauging at the level of the code. Along the way, we give CFT interpretations of the code subspace and the Hilbert space of qubits while mapping error operations to CFT defect fields.  
\end{abstract}
\maketitle

\section*{Introduction}

Quantum error correcting codes (QECCs) are integral to quantum computation. They also appear in high energy and condensed matter physics in various guises. As one important example, QECCs capture aspects of bulk reconstruction in AdS-CFT \cite{QECCAdSCFT}. Another notable case of a QECC in physics is the Toric code, a well-known model with topological order \cite{KitaevToric}. QECCs have also unravelled the existence and properties of fractons \cite{fractonsHaah}. More recently, QECCs were used to construct closed, simply connected manifolds \cite{manifoldsQECC}. 

In this work, we explore the relationship between conformal field theories (CFTs) in two spacetime dimensions, associated 3d Chern-Simons (CS) theories, and QECCs. The relationship between classical codes, their associated lattices, and holomorphic CFTs was originally noted by Dolan, Goddard, and Montague \cite{classCFT}. Recently, a quantum version of this relationship was discovered, where quantum stabilizer codes were associated with certain Narain rational CFTs (RCFTs) \cite{dymarsky2021quantum,dymarsky2021quantum2}. This construction does not exhaust all Narain RCFTs and leads to several natural questions: (1) When do general Narain RCFTs admit a quantum code description? (2) How does one identify the $n$-qubit Hilbert space, the code subspace and its complement, within the CFT Hilbert space? (3) What is the physical meaning of this relation?

In this work we answer these questions using the general structure of Narain RCFTs.\footnote{In principle, our results apply to any RCFT with abelian fusion rules (what we call an \lq\lq abelian RCFT") whether it admits a Narain description or not. In what follows, we will not attempt to distinguish between Narain RCFTs and hypothetically more general abelian RCFTs.} Our main results are:
\begin{itemize}
\item Any abelian CS theory with an even-order fusion group is related to a Narain RCFT that admits a stabilizer code description. Orbifolding this RCFT by a chiral algebra-preserving $Q\simeq\mathbb{Z}_2^k$ 0-form gauge group results in a Narain RCFT that continues to admit a stabilizer code description whenever the corresponding 3d bulk 1-form symmetry of the CS theory has vanishing 't Hooft anomaly.
\item All Narain RCFTs have abelian 0-form symmetries implemented by topological defects. In the class of theories described in the previous bullet, topological defect endpoint operators can naturally be mapped to the full Pauli group. The stabilizer subgroup corresponds to genuine local CFT operators, which can be thought of as living at the end of the trivial defect.
\item Under this map, the RCFT Hilbert space corresponds to the code subspace and certain defect Hilbert spaces correspond to the complement of the code subspace inside the $n$-qubit Hilbert space.
\end{itemize}

This paper is organized as follows. In Section I, we start with a brief review of stabilizer codes and Narain CFTs. We then show that Narain RCFTs with left and right movers paired via charge conjugation can be naturally associated with quantum stabilizer codes. We end Section I by extending this map to orbifold theories and deriving a relationship between vanishing 't Hooft anomalies and stabilizer codes; along the way, we consider various illustrative examples. In Section II, we study symmetries of Narain CFTs and show that operators living at the ends of topological defect lines implementing these symmetries give rise to the full Pauli group. We introduce the notion of a Verlinde subgroup and discuss its role in determining the error detection capability of CFT symmetry currents. In Section III, we propose a map between the $n$-qubit Hilbert space and states in the CFT. We conclude with a discussion and future directions.

\section{The Stabilizer Code / abelian RCFT Map}\label{ScRCFT}
Let us briefly review the basics of stabilizer codes and RCFTs with abelian fusion rules. We then propose a natural map relating them. 

A stabilizer code on $n$ qubits is defined by an abelian subgroup, $\CS_n$, of the generalized Pauli group on $n$ qubits, $\mathcal{P}_n$. Elements of $\mathcal{P}_n$ are defined by $\vec\alpha,\vec\beta\in\mathbb{Z}_2^n$ via
\bea\label{Gdef}
G(\vec \alpha,\vec \beta)&:=&X^{\alpha_1} \otimes \cdots \otimes X^{\alpha_n} \circ Z^{\beta_1} \otimes \cdots \otimes Z^{\beta_n}\cr&=& X^{\vec\alpha}\circ Z^{\vec\beta}\in\mathcal{P}_n~,
\eea
where the $i^{\rm th}$ $X$ and $Z$ are the Pauli matrices acting on the $i^{\rm th}$ qubit. This group has order $4^n$ and is non-abelian
\bea
\label{Paulicommutconst}
G(\vec \alpha_1,\vec \beta_1) G(\vec \alpha_2,\vec \beta_2)=(-1)^\epsilon ~G(\vec \alpha_2,\vec \beta_2) G(\vec \alpha_1,\vec \beta_1)~,
\eea
where $\epsilon(\vec\alpha_1,\vec\beta_1,\vec\alpha_2,\vec\beta_2):= {\vec\beta_1\cdot \vec\alpha_2-\vec\alpha_1\cdot \vec\beta_2}$.
The hallmark of a stabilizer subgroup is that any two elements commute with each other. Clearly, if $G(\vec \alpha_1,\vec \beta_1),G(\vec \alpha_2,\vec \beta_2)\in\CS_n$, then $G(\vec \alpha_1+\vec \alpha_2,\vec \beta_1+\vec \beta_2)\in\CS_n$. In this sense, stabilizer codes are additive. Moreover, all elements satisfy $G(\vec \alpha_i,\vec \beta_i)^2=1$. The states in the $n$-qubit Hilbert space  which are left invariant by  all $G\in \CS_n$ (i.e., $G\psi=\psi$)  are special: they form the \lq\lq code subspace." 
 
 The refined enumerator polynomial (REP) of an $n$ qubit stabilizer code is defined as 
\be
\label{REP}
W(x_1,x_2,x_3,x_4):= \sum_{G\in \CS_n} x_1^{\omega_I} x_2^{\omega_X} x_3^{\omega_Y} x_4^{\omega_Z}~,
\ee
where $\omega_{I/X/Y/Z}(G)$ count the number of $I/X/Y/Z$ Pauli matrices in the stabilizer group element $G$. 

For our general construction below, it is useful to  keep in mind  that the description above contains redundancies. In particular, two stabilizer codes are physically equivalent if they are related by an action of the Clifford group -- an outer automorphism of the Pauli group \cite{Gottesman:1997qd}. This group includes all $3!$ permutations of Pauli generators acting on each qubit.

The  stabilizer codes that play a role in \cite{dymarsky2021quantum} are self-dual: in other words $|\CS_n|=2^n$, and so there is a one-dimensional code subspace. These codes are also real (in the sense that all elements of $\CS_n$ in the representation \eqref{Gdef} are real-valued), but we will relax this latter condition in our general construction.   In the conventions of this paper, the map between the CFTs  and stabilizer codes  introduced in \cite{dymarsky2021quantum} is related to our map by an $X\leftrightarrow Y$ code equivalence.

The mapping between stabilizer codes and CFTs  associates classes of CFT operators with elements of $\CS_n$. Since the code is additive, we consider CFTs with additive (abelian) fusion rules (i.e., those corresponding to a lattice)
\begin{equation}\label{CFTfusion}
\phi_{\vec P_L,\vec P_R}\times\phi_{\vec K_L,\vec K_R}=\phi_{\vec P_L+\vec K_L,\vec P_R+\vec K_R}~,
\end{equation}
where the pair of vector indexes label left-moving and right-moving momenta valued in a Narain lattice, $\Lambda$. We will use the terms \lq\lq Narain theories" and \lq\lq abelian CFTs" interchangeably. Since there are infinitely many CFT operators and finitely many elements of $\CS_n$, we must organize the CFT operators into finitely many equivalence classes. In the context of abelian RCFT, this naturally happens since each $\phi_{\vec P_L,\vec P_R}$ in \eqref{CFTfusion} satisfies
\begin{equation}
\phi_{\vec P_L,\vec P_R}\in (N_L,N_R)~,\ N_L\in {\rm Rep}(V_{L})~,\ N_R\in {\rm Rep}(V_{R})~,
\end{equation}
where $N_{L}$ ($N_R$) are one of finitely many representations of the left (right) moving chiral algebra, $V_L$ ($V_R$). For simplicity, we will only consider CFTs with $V_L=V_R=V$. 

Specializing to $V_L=V_R=V$ and satisfying some additional mild assumptions detailed in \cite{Frohlich:2009gb}, it turns out that any RCFT is a (generalized) orbifold of the \lq\lq Cardy case" RCFT for $V$. This latter RCFT, $\CT$, consists of operators built by pairing left and right movers transforming in ${\rm Rep}(V)$ that are related by charge conjugation.\footnote{Given $V$, it turns out that the charge-conjugation modular invariant CFT exists on very general grounds \cite{davydov2016}.} In the case of an abelian RCFT, the orbifold is a standard group orbifold of $\CT$ \cite{Fuchs:2004dz}. The $\CT$ RCFT is sometimes referred to as the \lq\lq charge conjugation modular invariant," and it has torus partition function\footnote{Note that the construction in \cite{Fuchs:2004dz} takes as input left and right moving chiral algebras and produces an RCFT valid on any genus surface.}
\begin{equation}\label{ccZ}
Z_{\CT}(q)=\sum_{\vec p}\chi_{\vec p}(q)\bar\chi_{\overline{\vec p}}(\bar q)~, \ \ \vec p+\overline{\vec p}=\vec 0~, \ \ N_{\overline{\vec p}},N_{\vec p},N_{\vec 0}\in{\rm Rep}(V_{\CT})~.
\end{equation}
{\it Here $\vec p$ is a vector labeling elements of ${\rm Rep}(V)$ (not an element of $\Lambda$} ),\footnote{We use capital $\vec P$ to denote lattice momentum and lower case $\vec p$ to denote elements of ${\rm Rep(V)}$.} we sum over characters describing the operator content of the theory, and $\overline{\vec p}$ labels the representation conjugate to $\vec p$.\footnote{This latter statement means that we have fusion of the form $N_{\vec p}\times N_{\overline{\vec p}}=N_{\vec 0}$.}

Mathematically, ${\rm Rep}(V)$ corresponds to a modular tensor category (MTC). Physically, ${\rm Rep}(V)$ labels Wilson lines in the 3d Chern-Simons (CS) theory related to the 2d RCFT in question (see Fig. 1). The full set of MTCs/CS theories related to our abelian RCFTs have been classified in \cite{Wall} (see also \cite{Wang:2020nmz}). The result is that any such CS theory is a direct product of arbitrary combinations of the following factors
\begin{eqnarray}\label{MTCclass}
A_{2^r}&\sim&\mathbb{Z}_{2^r}~,\ A_{q^r}\sim\mathbb{Z}_{p^r}~,\ B_{2^r}\sim\mathbb{Z}_{2^r}~,\cr B_{q^r}&\sim&\mathbb{Z}_{q^r}~,\ C_{2^r}\sim\mathbb{Z}_{2^r}~, \ D_{2^r}\sim\mathbb{Z}_{2^r}\ ,\cr E_{2^r}&\sim&\mathbb{Z}_{2^r}\times\mathbb{Z}_{2^r},\ \ \, F_{2^r}\sim\mathbb{Z}_{2^r}\times\mathbb{Z}_{2^r}~,
\end{eqnarray}
where the labels on the lefthand sides of \eqref{MTCclass} denote CS theories as in \cite{Wang:2020nmz} with fusion rules for Wilson lines given by the abelian groups on the righthand sides, and $q$ is an odd prime number.\footnote{Strictly speaking, since a given label on the lefthand side of \eqref{MTCclass} only specifies the statistics of a set of line operators, it can correspond to different CS theories. Moreover, a CS theory that does not factorize in the geometry with boundaries depicted in Fig. 1 with $M$ trivial can correspond to a product of labels (e.g., $U(1)_6$ CS theory, which corresponds to $B_2\times B_3$). For simplicity in what follows, we will avoid this latter possibility. \label{CSfactor}} The upshot is that we should think of $\vec p$ as valued in the following product group / lattice quotient
\begin{eqnarray}\label{pval}
\vec{p}\in&\prod_r&\Big(\mathbb{Z}_{2^r}^{n_{A_{2^r}}}\times\mathbb{Z}_{2^r}^{n_{B_{2^r}}}\times\mathbb{Z}_{2^r}^{n_{C_{2^r}}}\times\mathbb{Z}_{2^r}^{n_{D_{2^r}}}\times \Big[\mathbb{Z}_{2^r}\times\mathbb{Z}_{2^r}\Big]^{n_{E_{2^r}}}\cr&\times&\Big[\mathbb{Z}_{2^r}\times\mathbb{Z}_{2^r}\Big]^{n_{F_{2^r}}}\times\prod_q\Big[\mathbb{Z}_{q^r}^{n_{A_{q^r}}}\times\mathbb{Z}_{q^r}^{n_{B_{q^r}}}\Big]\Big):=K~,
\end{eqnarray}
where $n_{X}$ is the number of {\it independent} factors of the CS theory $X$  corresponding to the CFT in \eqref{ccZ} (see Footnote \ref{CSfactor}).\footnote{Here we are thinking of $\mathbb{Z}_N$ as an additive subgroup of $\mathbb{Z}$ modulo $N$.} Physically, {\it $K$ is the 1-form symmetry group of the CS theory and the 0-form symmetry subgroup of the RCFT that commutes with the full left and right chiral algebras} (see Fig. 2).

\begin{figure}
\centering

\tikzset{every picture/.style={line width=0.75pt}} 

\begin{tikzpicture}[x=0.55pt,y=0.55pt,yscale=-1,xscale=1]

\draw   (444.58,203.85) -- (382.59,235) -- (381.04,83.79) -- (443.03,52.64) -- cycle ;
\draw  [color={rgb, 255:red, 72; green, 17; blue, 210 }  ,draw opacity=1 ] (353.82,203.85) -- (291.83,235) -- (290.28,83.79) -- (352.27,52.64) -- cycle ;
\draw   (265.95,203.85) -- (203.95,235) -- (202.4,83.79) -- (264.4,52.64) -- cycle ;
\draw    (234.17,143.82) -- (290.38,144.08) ;
\draw  [dash pattern={on 0.84pt off 2.51pt}]  (290.38,144.08) -- (322.05,143.82) ;
\draw    (322.05,143.82) -- (381.14,144.08) ;
\draw  [dash pattern={on 0.84pt off 2.51pt}]  (381.14,144.08) -- (412.81,143.82) ;
\draw  [fill={rgb, 255:red, 0; green, 0; blue, 0 }  ,fill opacity=1 ] (316.29,143.82) .. controls (316.29,142.25) and (317.58,140.98) .. (319.17,140.98) .. controls (320.76,140.98) and (322.05,142.25) .. (322.05,143.82) .. controls (322.05,145.39) and (320.76,146.66) .. (319.17,146.66) .. controls (317.58,146.66) and (316.29,145.39) .. (316.29,143.82) -- cycle ;
\draw  [fill={rgb, 255:red, 0; green, 0; blue, 0 }  ,fill opacity=1 ] (228.41,143.82) .. controls (228.41,142.25) and (229.7,140.98) .. (231.29,140.98) .. controls (232.88,140.98) and (234.17,142.25) .. (234.17,143.82) .. controls (234.17,145.39) and (232.88,146.66) .. (231.29,146.66) .. controls (229.7,146.66) and (228.41,145.39) .. (228.41,143.82) -- cycle ;
\draw  [fill={rgb, 255:red, 0; green, 0; blue, 0 }  ,fill opacity=1 ] (411.37,143.82) .. controls (411.37,142.25) and (412.66,140.98) .. (414.25,140.98) .. controls (415.84,140.98) and (417.13,142.25) .. (417.13,143.82) .. controls (417.13,145.39) and (415.84,146.66) .. (414.25,146.66) .. controls (412.66,146.66) and (411.37,145.39) .. (411.37,143.82) -- cycle ;
\draw    (227.69,264.5) -- (408.58,263.51) ;
\draw [shift={(410.58,263.5)}, rotate = 179.69] [color={rgb, 255:red, 0; green, 0; blue, 0 }  ][line width=0.75]    (10.93,-3.29) .. controls (6.95,-1.4) and (3.31,-0.3) .. (0,0) .. controls (3.31,0.3) and (6.95,1.4) .. (10.93,3.29)   ;
\draw   (235.42,267.68) .. controls (232.16,266.01) and (228.88,265.02) .. (225.58,264.7) .. controls (228.9,264.36) and (232.22,263.33) .. (235.58,261.65) ;

\draw (238,120.4) node [anchor=north west][inner sep=0.75pt]    {$W_{\vec p}$};
\draw (385,120.4) node [anchor=north west][inner sep=0.75pt]    {$W_{\vec q}$};
\draw (307,152.4) node [anchor=north west][inner sep=0.75pt]    {$$};
\draw (316,35.4) node [anchor=north west][inner sep=0.75pt]    {$M$};
\draw (205.4,84.19) node [anchor=north west][inner sep=0.75pt]    {$\Sigma $};
\draw (384.04,84.19) node [anchor=north west][inner sep=0.75pt]    {$\Sigma'$};
\draw (313,269.4) node [anchor=north west][inner sep=0.75pt]    {$I$};

\end{tikzpicture}
\caption{The pairing of 2d CFT left and right movers on $\Sigma$ and $\Sigma'$ can be specified by an abelian CS theory on $X\simeq\Sigma \times I$ with a surface operator, $M$, inserted in between \cite{Kapustin:2010if,Komargodski:2020mxz}. A local operator, $O_{(\vec p,\vec q)}$, is specified by the Wilson lines $W_{\vec p}$ and $W_{\vec q}$. Different $M$ lead to different partition functions. Topological defects in the 2d CFT correspond to Wilson lines parallel to $\Sigma, \Sigma'$ (see Fig. 2).}
\end{figure}
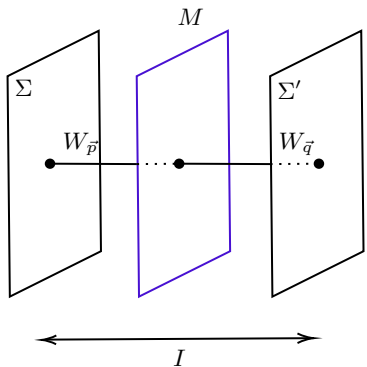

\begin{figure}
\centering

\tikzset{every picture/.style={line width=0.75pt}} 

\begin{tikzpicture}[x=0.55pt,y=0.55pt,yscale=-1,xscale=1]

\draw   (444.58,203.85) -- (382.59,235) -- (381.04,83.79) -- (443.03,52.64) -- cycle ;
\draw  [color={rgb, 255:red, 72; green, 17; blue, 210 }  ,draw opacity=1 ] (353.82,203.85) -- (291.83,235) -- (290.28,83.79) -- (352.27,52.64) -- cycle ;
\draw   (265.95,203.85) -- (203.95,235) -- (202.4,83.79) -- (264.4,52.64) -- cycle ;
\draw    (227.69,264.5) -- (408.58,263.51) ;
\draw [shift={(410.58,263.5)}, rotate = 179.69] [color={rgb, 255:red, 0; green, 0; blue, 0 }  ][line width=0.75]    (10.93,-3.29) .. controls (6.95,-1.4) and (3.31,-0.3) .. (0,0) .. controls (3.31,0.3) and (6.95,1.4) .. (10.93,3.29)   ;
\draw   (235.42,267.68) .. controls (232.16,266.01) and (228.88,265.02) .. (225.58,264.7) .. controls (228.9,264.36) and (232.22,263.33) .. (235.58,261.65) ;
\draw [color={rgb, 255:red, 208; green, 2; blue, 27 }  ,draw opacity=1 ]   (276.4,58) .. controls (276.4,133) and (284.4,169) .. (239.4,174) ;
\draw  [fill={rgb, 255:red, 0; green, 0; blue, 0 }  ,fill opacity=1 ] (233.64,174) .. controls (233.64,172.43) and (234.93,171.16) .. (236.52,171.16) .. controls (238.11,171.16) and (239.4,172.43) .. (239.4,174) .. controls (239.4,175.57) and (238.11,176.84) .. (236.52,176.84) .. controls (234.93,176.84) and (233.64,175.57) .. (233.64,174) -- cycle ;

\draw (324,28.4) node [anchor=north west][inner sep=0.75pt]    {$M$};
\draw (205.4,84.19) node [anchor=north west][inner sep=0.75pt]    {$\Sigma $};
\draw (385.04,84.19) node [anchor=north west][inner sep=0.75pt]    {$\Sigma '$};
\draw (313,269.4) node [anchor=north west][inner sep=0.75pt]    {$I$};
\draw (268,38.4) node [anchor=north west][inner sep=0.75pt]    {$\mathcal{L}_{\vec p}$};

\end{tikzpicture}
\caption{The endpoint of $\CL_{\vec p}$ on $\Sigma$ gives a defect endpoint operator corresponding to a state in the defect Hilbert space, $\CH_{\CL_{\vec p}}^{\rm Defect}$. We can think of $\CL_{\vec p}$ as generating a 3d 1-form symmetry or a 2d 0-form symmetry (when $\CL_{\vec p}$ is pushed to lie completely on $\Sigma$).}
\end{figure}
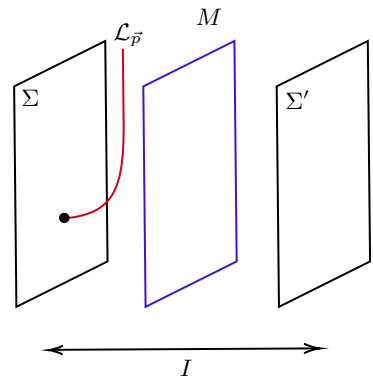

Now we will map the pair $(\vec\alpha,\vec\beta)$, which  specifies a stabilizer generator from $\CS_n$ to a pair $(\vec p,\overline{\vec p})$ representing a family of operators contributing to $\chi_{\vec p}\bar\chi_{\overline{\vec{p}}}$ in \eqref{ccZ}. First we  specify the dimension of $\vec p$: the most obvious choice is that $\vec\alpha,\vec\beta$, and $\vec p$ are $n$-dimensional. Moreover, in our map $\vec\alpha$ and $\vec\beta$ are linearly related to $\vec p$.

To begin with, let us neglect possible $E_{2^r}$ and $F_{2^r}$ CS theory factors. Then, $\CT$ is a CFT with $n$ decoupled factors having fusion rules given by the $n$ factors in \eqref{pval}.\footnote{More explicitly, we have that \\ $n=\sum_r\left(n_{A_{2^r}}+n_{B_{2^r}}+n_{C_{2^r}}+n_{D_{2^r}}+\sum_{q}\left(n_{A_{q^r}}+n_{B_{q^r}}\right)\right)$.} Indeed, by construction, each of the $n$ CFT factors is closed under fusion.\footnote{If we relax the condition in Footnote \ref{CSfactor} and allow for CS theories like $U(1)_6$, then we can also consider charge conjugation modular invariants that do not decompose into $n$ such CFT factors.} It is therefore natural to associate such a theory with an $n$-fold product of one-qubit codes. Up to code equivalence, all such codes are generated by $Z$ acting on individual qubits. Therefore, we set $\vec\alpha=0$, and choose
\begin{equation}\label{linmap}
\vec\beta=\vec p~,
\end{equation}
where \eqref{linmap} is the simplest natural choice.

However, note that for a CFT factor described by $A_{q^r}$ or $B_{q^r}$, the simplest choice is to make the resulting code factor trivial. The reason is that the corresponding component of $\vec p$, $p_i$, has order $q^s$ for  $1\le s\le r$. In this case, multiple stabilizers would correspond to the same $(\vec p,\bar{\vec p})$. We therefore ignore factors described by $A_{q^r}$ and $B_{q^r}$ from now on and map corresponding CFT degrees of freedom to 0-qubit codes.

In summary, we learn that linearity and code redefinitions point to the relation
\begin{equation}
\left\{\CO_{\vec p,\overline{\vec{p}}}\right\}\,\leftrightarrow\, Z^{\vec p}~,
\end{equation}
where we understand this map as meaning that the $Z^{\vec p}$ stabilizer corresponds to the collection of operators in the $(\vec p,\overline{\vec p})$ representation of the left and right moving chiral algebras (i.e., the primary and its descendants). Including factors of $E_{2^r}$ and $F_{2^r}$ and following logic similar to the above leads to the map
\begin{equation}\label{ccZrel}
\left\{\CO_{\vec p,\overline{\vec{p}}}\right\}\, \leftrightarrow\, Z^{A\vec p}~,
\end{equation}
where $A$ is block diagonal, with the following diagonal components corresponding to different CFT factors
\begin{equation}
A_{A_{2^r}}=A_{B_{2^r}}=A_{C_{2^r}}=A_{D_{2^r}}=1~,
\end{equation}
and, up to code equivalence, 
\begin{equation}\label{EFchoice}
A_{E_{2^r}}=A_{F_{2^r}}=\begin{pmatrix}
0 & 1\\
1 & 0
\end{pmatrix}~.
\end{equation}
Note that in writing \eqref{ccZrel}, we allow for multiple families of operators to appear on the lefthand side (see Section \ref{Examples1} for some examples). Indeed, the exponent of $Z$ on the RHS is only sensitive to $A\vec p$ modulo two. Thus in the simple case of charge conjugation modular invariant, we have the CFT to stabilizer code map
\begin{equation}\label{MuMap}
\mu:\CT\longrightarrow S_{\CT}:={\rm gen}\left\{Z^{A\vec e_i}\ |\ e_{ij}=\delta_{ij}\right\}~,
\end{equation}
where \lq\lq${\rm gen}\left\{\cdots\right\}$" means that the code is generated by the enclosed Pauli operators. Note that this code is self-dual by construction. Moreover, $\mu$ is non-invertible. For example, the $SU(2)$ and $E_7$ WZW models at level one are distinct but map to the same code.\footnote{The reason is that in both cases, $\vec p=p_1$ takes values in the same group.}

Given the set of theories of the form \eqref{ccZ}, we can construct all other Narain RCFTs by orbifolding them by some non-anomalous 0-form symmetry subgroup $Q\lhd K$.\footnote{As we will see, the theories in \cite{dymarsky2021quantum} are all orbifolds of particular theories with partition functions of the form \eqref{ccZ}. Note that we will only consider orbifolds with respect to symmetries which commute with the full left and right chiral algebras. Orbifolds of this type take us from a Narain CFT to another Narain CFT, while more general orbifolds may result in non-Narain CFTs.} Here non-anomalous means that the associator of Verlinde lines implementing $Q$ is trivial in $H^3(Q,U(1))$. \footnote{For the CFT with charge conjugation modular invariant, $F$ can be written in terms of holomorphic scaling dimensions as 
\be
F(\vec g, \vec h,\vec k)= \prod_i \bigg\{
	\begin{array}{ll}
		1 & \mbox{   if } h_i + k_i < n_i\\
		\theta(e_i)^{g_in_i}       & \mbox{  if } h_i + k_i \geq n_i 
	\end{array}
\ee
where $e_i$ is a basis for the cyclic factors in \eqref{pval}, and $\vec g= \sum_i g_i e_i$. Here $n_i$ is the order of the $i^{\text{th}}$ cyclic factor, and $\theta_{\vec p}:={\rm exp}(2\pi ih_{\vec p})$, where $h_{\vec p}$ is the holomorphic scaling dimension of an operator in representation $\vec p$. The group $Q$ is non-anomalous if and only if $\theta_{\vec h}^{O_{\vec h}}=1 ~ \forall \vec h \in Q$, where $O_{\vec h}$ is the order of $\vec h$ in $Q$ \cite{Fuchs:2004dz}.} Therefore, if $Q$ is non-anomalous, $F$ is a 3-coboundary satisfying
\be
F(\vec h_1,\vec h_2,\vec h_3)= \frac{\tau(\vec h_2, \vec h_3) \tau(\vec h_1,\vec h_2 + \vec h_3) }{\tau(\vec h_1 + \vec h_2, \vec h_3)   \tau(\vec h_1, \vec h_2) }   ~ \forall \vec h_1, \vec h_2, \vec h_3 \in Q~,
\ee
where $\tau$ is a 2-cochain.  
Then, the $Q$-orbifold torus partition function is
\begin{equation}
\label{HZ}
Z_{\CT/Q,[\sigma]}=\sum_{\vec g\in H}\sum_{\vec p\in B_{\vec g}}\chi_{\vec p}(q)\bar\chi_{\overline{\vec p+\vec g}}(\bar q)~,
\end{equation}
where $[\sigma]$ is an equivalence class in $H^2(Q,U(1))$ defining the discrete torsion (in the condensed matter perspective, the 2d SPT we stack when gauging $Q$, or the $B$-field in \cite{dymarsky2021quantum}), and 
\be
\label{pconst2}
B_{\vec g}:=\left\{\vec p\ \Big|\ S_{\vec h,\vec  p} ~ \Xi(\vec h,\vec g)=1 ~,\ \forall \vec h \in H\right\}~,
\ee
where we define\footnote{Note that our $S$ matrix differs from the unitary $S$ matrix by an overall normalization (ours is $\sqrt{N}$ times bigger, where $N$ is the number of Wilson lines in the CS theory associated with our RCFT)\label{normS}.} 
\bea
\label{SXiZ2k}
S_{\vec h,\vec p}:={\theta_{\vec h+\vec p}\over\theta_{\vec h}\theta_{\vec p}}~,\ \Xi(\vec g,\vec h)&:=&R(\vec{h},\vec g) \frac{\tau(\vec h,\vec g)\sigma(\vec h,\vec g)}{\tau(\vec g, \vec h)\sigma(\vec g, \vec h)}~.\ \ 
\eea
In \eqref{SXiZ2k}, $\theta_{\vec p}:={\rm exp}(2\pi ih_{\vec p})$, and $h_{\vec p}$ is the holomorphic scaling dimension of an operator in representation $\vec p$.\footnote{$R(\vec h, \vec g)$ can be written in terms of $\theta_{\vec g}$ as
$R(\vec h,\vec g)= \prod_i (\theta_{e_i})^{h_ig_i} \prod_{i<j} (S_{e_i,e_j})^{h_i g_j}$, where $e_i$ is a basis for the cyclic factors in \eqref{pval}, and $\vec g= \sum_i g_i e_i$. Note that both $R(\vec h, \vec g)$ and $\tau(\vec g, \vec h)$ depend on a choice of basis in Rep$(V)$, but $\Xi(\vec g, \vec h)$ is basis independent.}

In this paper we focus on the case
\begin{equation}\label{Horbgroup}
Q\simeq\mathbb{Z}_2^k~.
\end{equation}
Such subgroups are the most universal in the sense that they are contained in any other subgroups of $K$.\footnote{Recall that we are ignoring CFT factors involving primaries labeled by $A_{q^r}$ and $B_{q^r}$.} More general cases can be treated in a similar fashion.

How should we include the data of states corresponding to $\vec g\ne\vec0$ in the code? Clearly, the fields in the $\vec g=\vec 0$ sector should still be captured by \eqref{ccZrel}. Therefore, $\vec g$ must appear in a linear relation with $\vec \alpha,\vec\beta$ such that setting $\vec g=\vec0$ recovers terms of the form \eqref{ccZrel}. Note that nontrivial components of any $\vec g\in H$ have the form $g_i=2^{r_i-1}\in \mathbb{Z}_{2^{r_i}}$ (since $\vec g+\vec g=\vec 0$). Therefore, in order to contribute to the stabilizer, $\vec g$ must appear through $M\vec g$ ($M$ is diagonal, and $M_{ii}:=2^{1-r_i}$).

At this point, we should ask what principle requires $\vec g$ to contribute to the stabilizers at all. The answer is that orbifolding is an invertible procedure: when we gauge a discrete 0-form symmetry, $Q$, of a CFT, $\CT$,\footnote{Note that to unambiguously refer to the orbifolded theory, we should also generally specify the discrete torsion, $[\sigma]$. However, we will often be slightly imprecise and leave the discrete torsion implicit in our discussions.} there is an isomorphic dual $Q'\simeq Q$ symmetry we can gauge in $\CT/Q$ to return back to the original theory.\footnote{See \cite{Ginsparg:1988ui,DiFrancesco:1997nk} as well as the more recent discussion in \cite{Yuji}.} {\it We would like this invertibility to extend to the map between codes.}

If $M\vec g$ only appears through a factor $Z^{M\vec g}$, then our map between codes will not generally be invertible. The simplest and most natural possibility is the following.\footnote{We can also include an $M\vec g$ contribution in $Z$. Then we have $X^{M \vec g} \circ Z^{A \vec p + M \vec g}= Y^{M \vec g} \circ Z^{A \vec p }$ which is equivalent to the code $X^{M \vec g} \circ Z^{A \vec p }$. Similarly, $X^{M \vec g + A \vec p} \circ Z^{A \vec p}$ is code equivalent to $X^{M \vec g} \circ Z^{A \vec p}$.}

\noindent
{\bf CFT to stabilizer operator map:}
\be\label{codeZ}
\left\{\CO_{\vec p,\overline{\vec g+\vec p}}\right\}\, \leftrightarrow\,  X^{M \vec g} \circ Z^{A \vec p }:=G(M\vec g,A\vec p)~.
\ee
In the language of \eqref{MuMap}, we have
\begin{equation}\label{MuMap2}
\mu:\CT/Q\longrightarrow S_{\CT/Q}:={\rm gen}\left\{X^{M\vec g_i}Z^{A\vec p_J}\right\}~,
\end{equation}
where $\vec g_i$ and $\vec p_J$ generate $Q$ and $K$ respectively.

Since $Z$ is order two, the quantum code constructed above is only sensitive to $A \vec{p}_J$ mod 2. Therefore, in general we will have multiple families of operators mapping to the same element of the stabilizer group.

Recall that the stabilizer code associated with the charge conjugation modular invariant is self-dual. Since orbifolding is invertible, the above map assigns a self-dual code to $\CT/K$ (see Appendix B for an alternate argument).

Intriguingly, given the map in \eqref{codeZ}, the commutation relations of elements of $\CS_{\CT/H}$ are controlled by the $S$ matrix of the RCFT. Indeed, it is a simple exercise to check that
\bea\label{stabCond}
G(\vec g_{1},\vec p_{1}) G(\vec g_{2},\vec p_{2})&=&
e^{\pi i [M \vec g_{2} \cdot A \vec p_{1}- M \vec g_{1} \cdot A \vec p_{2}] }\cr&\cdot& \ \ G(\vec g_{2},\vec p_{2}) G(\vec g_{1},\vec p_{1}) \nonumber \\
&=& S_{\vec g_{2}, \vec p_{1} } S_{\vec g_{1}, \vec p_{2} }\cr&\cdot& \ \ G(\vec g_{2},\vec p_{2}) G(\vec g_{1},\vec p_{1})\nonumber\\
&=& \Xi(\vec g_{2}, \vec g_{1} ) \Xi(\vec g_{1}, \vec g_{2} )\cr&\cdot& \ \ G(\vec g_{2},\vec p_{2}) G(\vec g_{1},\vec p_{1})\nonumber \\
&=& S_{\vec g_{1}, \vec g_{2}} \cr&\cdot& \ \ G(\vec g_{2},\vec p_{2})G(\vec g_{1},\vec p_{1})~,
\eea
where, in the third equality, we have used \eqref{pconst2}. We have also used the expression for the $S$ matrix $S_{\vec p, \vec q}=e^{\frac{2 \pi i }{2} \vec p^T M A \vec q }$ which follows from \eqref{pval} \cite{Wang:2020nmz}. Therefore, $\CS_{\CT/H}$ is a stabilizer code if and only if $S_{\vec g_1, \vec g_2}=1$. This latter statement can be reinterpreted as the vanishing of the 1-form anomaly for the $Q$ 1-form symmetry in the bulk CS theory related to the $\CT$ RCFT.

\subsection{Examples}\label{Examples1}

\subsubsection{R=1,2 compact boson}

The code CFTs in \cite{dymarsky2021quantum} are all orbifolds of charge conjugation modular invariants with Rep$(V)=A_4^{n_{A_4}}$, for some integer $n_{A_4}>0$. That is, the fusion rules for the charge conjugation modular invariants are given by the abelian group, $K=(\DZ_4)^{n_{A_4}}$ (all other $n_X$ in \eqref{pval} vanish). The theories discussed in \cite{dymarsky2021quantum} with non-trivial $B$-field correspond in our language to orbifolds of the charge conjugation theories with discrete torsion turned on (or, equivalently, a non-trivial 2D SPT in the $\mathbb{Z}_2^k\lhd \DZ_4^{n_{A_4}}$ gauging process). As such, the CFTs in \cite{dymarsky2021quantum} are a small subset of theories discussed here.

The simplest code CFT among these is the $R=1$ compact boson, corresponding to the choice $n_{A_4}=1$. The chiral algebra has the trivial, fundamental, spinor, and conjugate spinor representations which we will denote by $N_{0},N_{2},N_{1},N_{3}$, respectively. These form the $K=\DZ_4$ group under fusion. The scaling dimensions of chiral primaries in these representations are
\be\label{chirscale}
h_{0}=0~,\ h_{2}=\frac{1}{2}~,\ h_{1}=h_{3}=\frac{1}{8}~.
\ee
The Narain lattice for this theory is given by
\bea
\label{R=1Narain}
 P_L := n+\frac{m}{2}~, ~ P_R := n-\frac{m}{2}~,
\eea
where $m,n\in \DZ$. In general, the vertex operators are given by 
\be
\label{vertexop}
V_{(n,m)}= : e^{i \vec p_L \vec X_L} e^{i \vec p_R \vec X_R}:~,
\ee
where $\vec X_L,\vec X_R$ are the left and right moving components of the field $X$ describing the compact boson. 
The partition function is
\be
Z_{\CT}=\chi_0\bar\chi_0+\chi_{2}\bar\chi_{2}+\chi_1\bar\chi_3+\chi_{3}\bar\chi_{ 1}~,
\ee
which is the charge conjugation modular invariant. The scaling dimensions of the primaries are twice those in \eqref{chirscale}. Here $\chi_{i}$ is the character of $N_{i}$ given by \cite{DiFrancesco:1997nk}
\be
\chi_{p}(q)=\frac{1}{\eta(q)}\sum_{n \in \DZ} q^{2 (n+\frac{p}{4})^2}~,
\ee
where $p=0,1,2,3$ and $\eta$ is the Dedekind eta function. 
Note that the partition function can also be written in terms of the Narain lattice vectors as
\be
Z_{\CT}(\tau, \bar \tau)= \frac{1}{|\eta(\tau)|^2}\sum_{(P_L,P_R)} q^{\frac{P_L^2}{2}} \bar q^{\frac{P_R^2}{2}}, ~ q=e^{2\pi i \tau}, ~ \bar q=e^{-2\pi i \bar \tau}  
\ee
The lattice vectors corresponding to a primary operator $\CO_{p,\bar p}$ can be found by requiring 
\be
\frac{P_L^2 + P_R^2}{2}=2 h_{\vec p} 
\ee
where the R.H.S. is the scaling dimension of $\CO_{p,\bar p}$. In particular, the primary operators $\mathcal{O}_{1,3}, \mathcal{O}_{3,1}$ correspond to the lattice vectors
\be
(P_L,P_R)=\bigg (\frac{1}{2},-\frac{1}{2} \bigg )~ , \bigg (-\frac{1}{2},\frac{1}{2}\bigg )~,
\ee
while $\mathcal{O}_{2,2}$ corresponds to\footnote{The four states in \eqref{22degen} correspond to the fact that $\CO_{2,2}$ transforms as a left-moving $so(2)$ vector times a right moving $so(2)$ vector.}
\be\label{22degen}
(P_L,P_R)=(1,1) \oplus (1,-1) \oplus (-1,1) \oplus (-1,-1)~,
\ee
and $\CO_{0,0}=\mathds{1}$ to (0,0). We can assign each $(P_L,P_R)$ lattice point to be in a particular $\left\{\CO_{p,\bar p}\right\}$ family by considering fusions of the above operators and imposing that fusions correspond to momentum vector addition. Using \eqref{ccZrel}, these operators map to the 1-qubit stabilizer code generated by the $Z$ Pauli matrix via 
\be
I \leftrightarrow \{\mathcal{O}_{0,0}\},\{\mathcal{O}_{2,2}\}~,~ Z \leftrightarrow \{\mathcal{O}_{1,3}\},\{\mathcal{O}_{3,1}\}~,
\ee
where the map includes all descendants.

A topological line operator, denoted $\CL_{2}$, labelled by $\vec p=2$ generates a $\DZ_2$ 0-form symmetry. This symmetry acts by a shift $\phi \rightarrow \phi - \pi$, where $\phi:=\frac{X_L - X_R}{2}$. The action on the vertex operators is
\be
V_{(n,m)} \rightarrow (-1)^m V_{(n,m)}
\ee
In particular, the collections of operators $\{\mathcal{O}_{1,3}\}, \{\mathcal{O}_{3,1}\}$ change sign under this symmetry while $\{\mathcal{O}_{0,0}\}, \{\mathcal{O}_{2,2}\}$ remain invariant. 
This symmetry is non-anomalous because $h_{2}=\frac{1}{2}$ \cite{Fuchs:2004dz} (see also the related discussion in \cite{Lin:2019kpn} and Footnote 12). Taking the $\DZ_2$-orbifold,\footnote{$H^2(\DZ_2,U(1))\cong \DZ_1$. Therefore, there is no discrete torsion.} we get a dual CFT with partition function (using \eqref{HZ}, \eqref{pconst2})
\be
Z_{\CT/\DZ_2}=\chi_0\bar\chi_0+\chi_{2}\bar\chi_{2}+\chi_1\bar\chi_1+\chi_{3}\bar\chi_{3}~.
\ee
This is the partition function of the $R=2$ compact boson, which is T-dual to the $R=1$ compact boson. Using \eqref{codeZ}, the stabilizer code corresponding to this CFT is the 1-qubit code generated by $Y$ via the map
\be
I \leftrightarrow \{\mathcal{O}_{0,0}\},\{\mathcal{O}_{2,2}\}; ~ Y \leftrightarrow \{\mathcal{O}_{1,1}\},\{\mathcal{O}_{3,3}\}~.
\ee
T-duality between these theories is captured by the fact that the 1-qubit code generated by $Y$ is equivalent to the code generated by $Z$ \cite{dymarsky2021quantum} (recall that our conventions here differ from those in \cite{dymarsky2021quantum} by an $X\leftrightarrow Y$ code equivalence). 

Using \eqref{REP}, we can compute the refined enumerator polynomials (REPs) for the codes above, generated by $Z$ and $Y$ to get 
\bea 
\label{ZREP}
W_{\text{gen}(Z)}(x_1,x_2,x_3,x_4)&=&x_1+x_4~, \nonumber\\
W_{\text{gen}(Y)}(x_1,x_2,x_3,x_4)&=&x_1+x_3~.
\eea
Therefore, corresponding CFT torus partition functions can be written in terms of the REPs by choosing
\bea
x_1&=& \chi_0\bar\chi_0+\chi_{2}\bar\chi_{2}~,\nonumber\\ x_4&=& \chi_1\bar\chi_3+\chi_{3}\bar\chi_{1}~,\nonumber\\ x_3&=& \chi_1\bar\chi_1+\chi_{3}\bar\chi_{3}~.
\eea

As a final note, let us comment that we obtain the same quantum codes using any RCFT with ${\rm Rep}(V)=A_4^{n_{A_4}}$. For any $n_{A_4}$ there are always infinitely many such RCFTs. For example, we can take the product of the $R=1$ compact boson with arbitrarily many $E_8$ WZW models at level one and trivial ${\rm Rep}(V)$ (this latter theory is associated with a $0$-qubit code). In this case, to get the partition function from the REP we have to input the characters $\chi_p \bar \chi_{\overline{\vec p}} \chi'_{0} \bar \chi'_{0}$ into \eqref{ZREP}, where $\chi'_{0}$ is the vacuum character of the $E_8$ WZW model at level 1 factors.

\subsubsection{$R=\sqrt{2}$ compact boson $\sim$ $SU(2)$ level one WZW} 

The compact boson at the self-dual radius, or, equivalently, the $SU(2)$ at level one WZW model has Rep$(V)=A_2$. That is, the representations of the chiral algebra are the trivial and fundamental representations, which we denote by $N_0,N_1$, respectively. They form a $K=\DZ_2$ group under fusion. We have chiral primaries with scaling dimensions
\be\label{chirSU(2)}
h_{0}=0~,\ h_{1}=\frac{1}{4}~.
\ee
The Narain lattice for this theory is given by
\bea
\label{latticesu2}
P_L := \frac{1}{\sqrt{2}} (n + m)~, ~ P_R := \frac{1}{\sqrt{2}} (n - m)~, 
\eea
where, $n,m\in \DZ$.
The vertex operators are given by \eqref{vertexop} with \eqref{latticesu2} inserted, and the torus partition function is
\be
Z_{\CT}=\chi_0\bar\chi_0+\chi_{1}\bar\chi_{1}~,
\ee
where the characters are given by \cite{DiFrancesco:1997nk}
\be
\chi_{p}(q)=\frac{1}{\eta(q)}\sum_{n \in \DZ} q^{\frac{(p+2n)^2}{4}}~,
\ee
with $i=0,1$.

The non-trivial primary, $\CO_{1,1}$, corresponds to the lattice vectors
\begin{equation}
(P_L,P_R)=\pm\left({1\over\sqrt{2}},{1\over\sqrt{2}}\right)\oplus\pm\left({1\over\sqrt{2}},-{1\over\sqrt{2}}\right)~,
\end{equation}
where the number of states follows from the fact that the primary transforms in the fundamental representation of the left and right moving $SU(2)$. We can assign any Narain lattice vector to be a member in a $\left\{\CO_{p,\bar p}\right\}$ family by considering fusions of the above primaries and imposing that they correspond to lattice vector addition. Now, using \eqref{ccZrel}, this CFT corresponds to the 1-qubit stabilizer code generated by $Z$ via the map
\be
I \leftrightarrow \{\mathcal{O}_{0,0}\}~, ~ Z \leftrightarrow \{\mathcal{O}_{1,1}\}~.
\ee
Note that this is the same quantum code as in the case of the $R=1$ compact boson. This fact illustrates that, the map \eqref{codeZ} can give the same quantum code for distinct CFTs.

The REP for this code is given by \eqref{ZREP},
and the torus partition function can be written in terms of $W$ by choosing
\be
x_1= \chi_0\bar \chi_0~,\quad  x_4= \chi_1\bar \chi_1~.
\ee

This CFT has a $\DZ_2$ 0-form symmetry generated by the topological line $\CL_2$. However, this $\DZ_2$ is anomalous \cite{Fuchs:2004dz} (see Footnote 12), and hence cannot be gauged (in a purely 2d system).

Again, from our construction, we can consider arbitrary products of this theory and, when we have at least two factors, orbifolds with and without discrete torsion.

\subsubsection{Compact boson at $R=\sqrt{\frac{2k}{\ell}}$}

Let us generalize the discussion above to compact boson at $R=\sqrt{\frac{2k}{\ell}}$, where $k,\ell$ are co-prime integers. This RCFT has fusion rules given by the group $K=\DZ_{2k\ell}$. The corresponding bulk CS theory is $U(1)_{2k\ell}$. Therefore, in this case Rep$(V)$ labels the Wilson lines in the $U(1)_{2k\ell}$ CS theory. Rep$(V)$ decomposes as follows 
\be
\label{Repdecomp}
\text{Rep}(V) \simeq X_{2^s} \times \prod_i (Y_i)_{q_i^{r_i}}~, K=\mathbb{Z}_{2^s}\times\prod_i\mathbb{Z}_{q_i^{r_i}}~.
\ee
where the $q_i$'s are distinct odd primes, $X\in \left\{A,B,C,D\right\}$, and $Y_i\in \left\{A,B\right\}$. Here the labels must be chosen so that the topological central charge is equal to $1$ modulo $8$. Note that this does not imply that the $U(1)_{2k\ell}$ CS theory or the associated CFT itself factorizes. The decomposition \eqref{Repdecomp} is an algebraic property of the set of representations of the chiral algebra Rep$(V)$ (see Footnote 6).

As discussed above, the odd factors contribute trivially to the code. For simplicity, we will therefore consider $\ell=2^{s-1}$ and $k=1$ for some integer $s>0$. In this case
\be
U(1)_{2^{s}} \text{ CS}\simeq A_{2^s}~, \ \ \ K=\mathbb{Z}_{2^s}~.
\ee
 The representations of the chiral algebra are denoted by integers $p \in \DZ_{2^{s}}$. The scaling dimensions for these chiral primaries are given by
$h_p= \frac{p^2}{2^{s+1}}$ if $p\le 2^{s-1}$ and $h_p= \frac{\bar p^2}{2^{s+1}}$ if $p>2^{s-1}$.

The Narain lattice for this theory is given by
\bea
\label{RNarain}
 P_L := \frac{n}{R}+\frac{mR}{2}~, ~ P_R := \frac{n}{R}-\frac{mR}{2}~,\ R=2^{2-s\over2}~,
\eea
where $m,n\in \DZ$. The vertex operators are given by \eqref{vertexop}.
 The torus partition function is
\be
Z_{\CT}= \sum_{p \in \DZ_{2^{s}}} \chi_{p} \bar \chi_{\bar p}~, 
\ee
which is the charge conjugation modular invariant. The characters, $\chi_p(q)$, are given by \cite{DiFrancesco:1997nk}
\be
\chi_p(q)= \frac{1}{\eta(g)} \sum_{n \in \DZ} q^{ 2^{s-1}\big (n+\frac{h_p}{2^{s}} \big )^2}~.
\ee

Non-trivial primaries, $\CO_{p,\bar p}$, with $p<2^{s-1}$ correspond to lattice vectors satisfying
\begin{equation}
{1\over2}(P_L^2+P_R^2)=2h_p~, P_L>P_R~,
\end{equation}
while the charge conjugate corresponds to lattice vectors of the above type with $P_R>P_L$. Finally, the non-trivial primary $\CO_{2^{s-1},2^{s-1}}$ corresponds to the lattice vectors satisfying
\begin{equation}
{1\over2}(P_L^2+P_R^2)=2^{s-2}~.
\end{equation}
The quantum code corresponding to this CFT is the 1-qubit quantum code generated by $Z$, where the operators are mapped to the code as 
\bea
I \leftrightarrow \{\CO_{p,\bar p}\} ~,~ p=\text{0 \text{ mod } 2}~, \cr
Z \leftrightarrow \{\CO_{p,\bar p}\} ~,~ p=\text{1 \text{ mod } 2}~.
\eea

A topological line operator, denoted $\CL_{2^{s-1}}$, labelled by $\vec p=2^{s-1}$ generates a $\DZ_2$ 0-form symmetry. This symmetry acts by a shift $\phi \rightarrow \phi - \pi$, where $\phi:=\frac{R(X_L - X_R)}{2}$. The action on the vertex operators is
\be
V_{(n,m)} \rightarrow (-1)^m V_{(n,m)}~.
\ee
This symmetry is non-anomalous and can be gauged. Taking the $\DZ_2$-orbifold we get the the orbifold CFT with the partition function
\be
Z_{\CT/\DZ_2}= \sum_{p= 0\text{ mod }2, p \in \DZ_{2^{s}}} \chi_{p} \bar \chi_{\bar p} + \chi_{p} \bar \chi_{\overline{2^{s-1} + p}}~, 
\ee
Using \eqref{codeZ}, the operators in this CFT can be mapped to the stabilizer code generated by $X$ as
\be
I \leftrightarrow \{\CO_{p,\bar p}\}, ~ X \leftrightarrow \{\CO_{p,\overline{2^{s-1}+p}}\}
\ee

Note that the quantum code corresponding to the $\DZ_2$ orbifold of the $R=1$ compact boson CFT is gen$(Y)$ while that for the $\DZ_2$ orbifold of the $R=2^{2-s\over2}$ compact boson CFT for $s>1$ is gen$(X)$. This difference is because, for $s>1$, the chiral primary $p=2^{s-1}$ is bosonic while, for $s=1$, it is fermionic. 

The REPs for the codes obtained above are 
\bea 
W_{\text{gen}(Z)}(x_1,x_2,x_3,x_4)=x_1+x_4~, \nonumber\\
W_{\text{gen}(X)}(x_1,x_2,x_3,x_4)=x_1+x_2~.
\eea
Therefore, the partition functions considered above can be written in terms of the REPs by choosing
\bea
x_1&=&  \sum_{p = 0\text{ mod }2} \chi_p \bar \chi_{\bar p}~, \nonumber\\ x_4&=& \sum_{p = 1\text{ mod }2} \chi_p \bar \chi_{\bar p}~,\nonumber\\ x_3&=& \sum_{p =0 \text{ mod }2} \chi_{p} \bar \chi_{\overline{2^{s-1} + p}}~.
\eea

\subsubsection{$\widehat{Spin(16)}_1$ CFT} 

The Spin$(16)_1$ CFT has Rep$(V)=E_2$ (the \lq\lq toric code" MTC). We denote the representations of the chiral algebra by $N_{(0,0)}, N_{(0,1)}, N_{(1,0)},N_{(1,1)}$, and they form a $K=\DZ_2 \times \DZ_2$ group under fusion. We have chiral primaries with scaling dimensions
\be
h_{(0,0)}=0~,\ h_{(0,1)}=h_{(1,0)}=1~, \ h_{(1,1)}={1\over2}~.
\ee
The Narain lattice is
\be
\{(\vec P_L,\vec P_R) \in \Lambda_{W} \times \Lambda_{W}, \vec P_L - \vec P_R \in \Lambda_R \}
\ee
where $\Lambda_W=\{\sum_{i} n_i \lambda_i, n_i \in \DZ\}$ is the weight lattice, $\lambda_i$ are the fundamental weights
\bea
\lambda_i= (1,\cdots,1,0,\cdots,0), ~ ~1\leq r\leq 6 \text{ (1 repeated $i$ times)}\nonumber \cr
\lambda_7= (1,1,1,1,1,1,1,1),~ \lambda_8=(1,1,1,1,1,1,1,-1)~,
\eea
and $\Lambda_R=\{\sum_{i} n_i \alpha_i, n_i \in \DZ\}$ where $\alpha_i$ are the simple roots
\be
\alpha_i= e_i - e_{i+1}~ 1 \leq i \leq 7, ~ \alpha_8=e_8+e_7~ .
\ee
Here $e_i$ is the vector with components $(e_i)_j=\delta_{i,j}$. It is easy to check that $\Lambda_R$ is the set of 8-component vectors such that the sum of its components is even. 

The partition function is
\be
Z_{\CT}=\chi_{(0,0)}\bar\chi_{(0,0)}+\chi_{(0,1)}\bar\chi_{(0,1)}+\chi_{(1,0)}\bar\chi_{(1,0)}+\chi_{(1,1)}\bar\chi_{(1,1)}~,
\ee
where the characters are given by \cite{DiFrancesco:1997nk}
\bea
\chi_{(0,0)}= \frac{(\theta_3^8+\theta_4^8)}{2\eta^8}&,& 
\chi_{(0,1)}=\chi_{(1,0)}=\frac{\theta_2^8}{2\eta^8}  ,\\
&&\chi_{(1,1)}= \frac{(\theta_3^8-\theta_4^8)}{2\eta^8} ~.
\eea
Here $\theta_2,\theta_{3},\theta_4$ are Jacobi-Theta functions. The Dynkin labels for the representations $N_{(0,0)},N_{(0,1)},N_{(1,0)}$ and $N_{(1,1)}$ are 
$(0,0,0,0,0,0,0,1), (0,0,0,0,0,0,1,0), (0,0,0,0,0,1,0,0)$ and $(1,0,0,0,0,0,0,0)$, respectively. Therefore, the primary operators $\CO_{(0,0),(0,0)},\CO_{(0,1),(0,1)},\CO_{(1,0),(1,0)}$ and $\CO_{(1,1),(1,1)}$, in turn, correspond to the lattice vectors
\bea 
(\lambda_8,\lambda_8)~,\ (\lambda_7,\lambda_7)~,\ (\lambda_6,\lambda_6)~,\ (\lambda_1,\lambda_1)~.
\eea
Using \eqref{ccZrel}, this CFT corresponds to the two-qubit stabilizer code generated by $I \otimes Z, Z \otimes I$ via the map
\be
I \otimes Z \leftrightarrow \{\mathcal{O}_{(1,0),(1,0)}\},~  Z \otimes I \leftrightarrow \{\mathcal{O}_{(0,1),(0,1)}\}~.
\ee

This CFT has three non-anomalous $\DZ_2$ 0-form symmetries, $Q_1,Q_2,Q_3$, corresponding to the topological lines $\CL_{(0,1)}, \CL_{(1,0)}$, and $\CL_{(1,1)}$. These symmetries act on the primary operators (and the corresponding Narain lattice vectors) as
\bea
\CL_{(0,1)}: \CO_{(1,0)} \rightarrow -\CO_{(1,0)}, ~ \CO_{(1,1)} \rightarrow -\CO_{(1,1)}~, \cr
\CL_{(1,0)}: \CO_{(0,1)} \rightarrow -\CO_{(0,1)}, ~ \CO_{(1,1)} \rightarrow -\CO_{(1,1)}~, \cr
\CL_{(1,1)}: \CO_{(0,1)} \rightarrow -\CO_{(0,1)}, ~ \CO_{(1,0)} \rightarrow -\CO_{(1,0)}~.
\eea
Actions of the symmetries on primaries not mentioned above are trivial. Orbifolding by $Q_1,Q_2,Q_3$, we get CFTs with partition functions (using \eqref{HZ}, \eqref{pconst2})
\bea
Z_{\CT/Q_1}= \chi_{(0,0)}\bar\chi_{(0,0)} &+& \chi_{(0,0)}\bar\chi_{(0,1)} + \chi_{(0,1)}\bar\chi_{(0,0)}\cr &&\ + \chi_{(0,1)}\bar\chi_{(0,1)} ~,\cr
Z_{\CT/Q_2}= \chi_{(0,0)}\bar\chi_{(0,0)} &+& \chi_{(0,0)}\bar\chi_{(1,0)} + \chi_{(1,0)}\bar\chi_{(0,0)}\cr &&\ + \chi_{(1,0)}\bar\chi_{(1,0)} ~,\cr  
Z_{\CT/Q_3}= \chi_{(0,0)}\bar\chi_{(0,0)} &+& \chi_{(1,1)}\bar\chi_{(1,1)} + \chi_{(0,1)}\bar\chi_{(1,0)}\cr&&\ + \chi_{(1,0)}\bar\chi_{(0,1)} ~,
\eea
respectively. Using \eqref{codeZ}, these CFTs can be mapped, in turn, to the stabilizer codes specified by gen$(Z \otimes I, I \otimes X)$, gen$(I \otimes Z, X \otimes I)$, and gen$(Z \otimes Z, Y \otimes X)$. 

We can also orbifold by the full $Q_1 \times Q_2$ symmetry of the CFT. We get partition functions (using \eqref{HZ}, \eqref{pconst2})
\bea
Z_{\CT/Q_1 \times Q_2, [1]}&=& \chi_{(0,0)}\bar\chi_{(0,0)} + \chi_{(0,1)}\bar\chi_{(0,0)} + \chi_{(0,0)}\bar\chi_{(1,0)} \nonumber \\
&& \hspace{3.0cm} + \chi_{(0,1)}\bar\chi_{(1,0)} ~,\cr
Z_{\CT/Q_1 \times Q_2, [\sigma]}&=& \chi_{(0,0)}\bar\chi_{(0,0)} + \chi_{(0,0)}\bar\chi_{(0,1)} + \chi_{(1,0)}\bar\chi_{(0,0)} \nonumber \\
&& \hspace{3.0cm}  + \chi_{(1,0)}\bar\chi_{(0,1)} ~,
\eea
where $[1]$ and $[\sigma]$ are the trivial and non-trivial elements of $H^2(\DZ_2 \times \DZ_2, U(1))$, respectively. Using \eqref{codeZ}, these CFTs can be mapped, in turn, to subgroups of the Pauli group specified by $\text{gen}(Z \otimes X, X \otimes I)$ and $\text{gen}(X \otimes Z, I \otimes X)$.

The subgroup of the Pauli group generated by these elements is clearly not a stabilizer code since it is non-abelian. For example, $Z \otimes X$ and $X \otimes I$ anti-commute with each other. This is expected from out general arguments above since $Q_1$ and $Q_2$ are related to 1-form symmetries of the bulk Spin$(16)_1$ Chern-Simons theory which have a mixed 't Hooft anomaly.

\section{Errors and the Full Pauli Group from Defects}\label{defects}
In the context of quantum codes, the elements of the Pauli group, $\mathcal{P}_n$, that are not in the stabilizer subgroup, $\CS_n$, are either called \lq\lq logical operators" or \lq\lq errors," depending, respectively, on whether they preserve the code subspace or map states from the code subspace to its complement. Since our codes are self-dual, we have no (non-trivial) logical operators,\footnote{Note that the elements in $\CS_n$, are sometimes called \lq\lq trivial" logical operators.} and all elements of $\mathcal{P}_n$ that are not in $\CS_n$ correspond to errors.

How can we see these errors in the CFT? An intuitive picture is provided by the toric code \cite{KitaevToric2}. There one finds that error operations correspond to string operators (defects) that create anyonic pairs.\footnote{For a pedagogical discussion, see section 11.3 of \cite{SteveSimon}.} When the anyons annihilate, the system returns to the code subspace, implementing a logical operation. While the gapped toric code system is very different from the CFTs considered in this paper, as we will see below, this geometric picture of errors is still informative. 

A more direct way to understand errors is to look  at the fields in $\CT/Q$ that contribute the terms with $\vec g\ne0$ in \eqref{HZ}. In the orbifolding procedure, we gauge $Q$ in the charge-conjugation modular invariant theory, $\CT$. The $\vec g\ne\vec 0$ bulk fields of $\CT/Q$ then come from certain fields living at the end of $Q$ topological defects of $\CT$. Therefore, the $X$-dependent Pauli stabilizers of the $\CT/Q$ theory appearing in \eqref{codeZ} correspond to error operations in the $\CT$ theory. This discussion suggests error operations of the code related to $\CT$ are given by defect endpoint operators of the $Q$ symmetries of $\CT$.  In the language of quantum codes, such orbifolding exchanges certain errors with stabilizers in an $n$-qubit self-dual code to produce a new $n$-qubit self-dual code, see e.g.~the examples in section \ref{defEx}. 

With the motivation above, we are now ready to identify the full set of error operations, i.e., to reconstruct the full Pauli group, from the defect fields. Since $Q$ consists of order-two defects which commute with the vacuum module, this suggests that we associate error operations with fields living at the ends of such defects. Through a slight abuse of terminology, we will refer to these and any other defects that preserve the maximal chiral algebra of a theory as \lq\lq Verlinde lines" (for further discussion of such lines, see for example \cite{Verlinde:1988sn,Petkova:2000dv,Gaiotto:2014lma,Buican:2017rxc,Chang:2018iay}).

To understand the spectrum of defect endpoint fields in the most general case, we eventually want to consider CFTs in which the pairing of characters is given by
\begin{eqnarray}\label{MZ}
Z_{\CT_{\CM}}=\sum_{\vec p,\vec q} \CM_{\vec p\vec q}\chi_{\vec p}(q)\bar\chi_{\vec q}(\bar q)~,
\end{eqnarray}
where $\CM$ is a matrix commuting with $S$ and $T$.\footnote{Here, we have $T_{\vec p,\vec q}:=e^{-\pi i(c/12)}\theta_{\vec p}\delta_{\vec p\vec q}$.} As a technically simpler starting point, let us first consider the case when $\CM_{\vec p\vec q}$ is a permutation on the set of vectors. Such modular invariants are called \lq\lq permutation modular invariants," and charge conjugation corresponds to the case $\CM_{\vec p,\vec q}=\delta_{\vec p,\bar{\vec q}}$. To avoid confusion below, we call theories of this type \lq\lq maximal" permutation modular invariants (MPMIs).\footnote{More general permutation modular invariants will play a role below.}  As we will see, we can reconstruct the Pauli group from Verlinde lines alone in any MPMI admitting a code description. 

In MPMIs, we define Verlinde lines via\footnote{In \eqref{VL} and bellow, $\vec\ell_{\CM}=\vec k$ is the unique vector such that $\CM_{\vec\ell\vec k}\ne0$.}
\begin{equation}\label{VL}
\CL_{(\vec p,\vec p_{\CM})}=\sum_{\vec\ell} {\bar S_{\vec p\vec\ell}\over \bar S_{\vec 0\vec\ell}}|\vec\ell,\vec\ell_{\CM}\rangle\langle\vec\ell,\vec\ell_{\CM}|~,
\end{equation}
where each $|\vec\ell,\vec\ell_{\CM}\rangle\langle\vec\ell,\vec\ell_{\CM}|$ is a projector on the  primary state labeled by $(\vec\ell,\vec\ell_{\CM})$ together with its descendants. Since this operator is a multiple of the identity within each representation of the left and right chiral algebras, it commutes with the chiral algebras and is topological (by construction, it commutes with the Virasoro sub-algebras). For convenience, we denote 
$\CL_{(\vec p,\vec p_{\CM})}$  simply as $\CL_{\vec p}$
since the right-moving label is determined by $\vec p$. Using the Verlinde formula, it is easy to check that these lines satisfy the fusion rules of the RCFT
\begin{equation}\label{Vfusion}
\CL_{\vec p}\times\CL_{\vec q}=\CL_{\vec p+\vec q}~.
\end{equation}
When $\vec p$ is order two, we have 
\begin{equation}\label{ord2}
\vec p+\vec p=\vec 0\ \Rightarrow\ S_{\vec p\vec\ell}/S_{\vec 0\vec\ell}\in\left\{\pm1\right\}~.
\end{equation}

To proceed, we insert $\CL_{\vec g}$ in the torus partition function (i.e., we wrap it on the spatial cycle of the torus) and perform a modular transformation so that it wraps time
\begin{eqnarray}\label{modTransD}
Z_{\CT_{\CM}}(\CL_{\vec \ell})&=&\sum_{\vec p,\vec q} {\bar S_{\vec\ell\vec p}\over \bar S_{\vec 0\vec\ell}}\CM_{\vec p\vec q}\chi_{\vec p}(q)\bar\chi_{\vec q}(\bar q)\cr&\to&\sum_{\vec p,\vec q,\vec r, \vec s} {\bar S_{\vec\ell\vec p}\over \bar S_{\vec 0\vec\ell}}\CM_{\vec p\vec q}S_{\vec p\vec r}\bar S_{\vec q\vec s}\chi_{\vec r}(q)\bar\chi_{\vec s}(\bar q)\cr&=&\sum_{\vec q,\vec r, \vec s} N_{\vec r\bar{\vec q}}^{\vec \ell} \CM_{\vec q\vec s}\chi_{\vec r}(q)\bar\chi_{\vec s}(\bar q):= Z_{\mathcal{T}_{\CM}}^{\vec\ell}(q,\bar q)~,\ \ \ \ \ \
\end{eqnarray}
where, in the last line, we have arrived at a definition for the partition function of fields living at the end of the defect labeled by $\vec\ell$. In the second to last equality, we use the Verlinde formula. In light of \eqref{Vfusion}, we can simplify the fusion coefficients as $N_{\vec r\bar{\vec q}}^{\vec\ell}=\delta_{\vec r-\vec q}^{\vec\ell}$. Therefore, we have
\begin{eqnarray}\label{permD}
Z_{\mathcal{T}_{\CM}}^{\vec\ell}(q,\bar q)&=&\sum_{\vec r,\vec s}\CM_{\vec r-\vec\ell\ \vec s}\chi_{\vec r}(q)\bar\chi_{\vec s}(\bar q)\cr&=&\sum_{\vec p,\vec q}\CM_{\vec p \vec q}\chi_{\vec p+\vec\ell}(q)\bar\chi_{\vec q}(\bar q)~.
\end{eqnarray}

Specializing to the case of the charge conjugation modular invariant, we obtain
\begin{equation}
Z_{\mathcal{T}}^{\vec\ell}(q,\bar q)=\sum_{\vec p}\chi_{\vec p+\vec\ell}(q)\bar\chi_{\bar{\vec p}}(\bar q)~.
\end{equation}
When $\vec\ell\in Q\simeq\mathbb{Z}_2^k$, we get, using $2\vec\ell=\vec0$,
\begin{eqnarray}
\label{chargedef}
Z_{\mathcal{T}}^{\vec\ell}(q,\bar q)&=&\sum_{\vec p}\chi_{\vec p+\vec\ell}(q)\bar\chi_{\overline{\vec\ell+\vec p+\vec\ell}}(\bar q)~.
\end{eqnarray}
As expected, these are equivalent to the contributions in \eqref{codeZ}, only here they correspond to defect operators in $\CT$ rather than bulk operators in $\CT/Q$. Therefore, consistency with the map in \eqref{codeZ} demands
\be\label{codeDZ}
\left\{\CO^{\vec\ell}_{\vec p+ \vec \ell,\overline{\vec p}}\right\}\, \leftrightarrow\,  X^{M \vec \ell} \circ Z^{A (\vec p+\vec \ell) }~,
\ee
where $\left\{\CO^{\vec\ell}_{\vec p+ \vec \ell,\overline{\vec p}}\right\}$ should be understood as an $\vec\ell$-defect primary operator and its associated descendants. If $Q\simeq\mathbb{Z}_2^n$, then \eqref{codeDZ} gives rise to the full Pauli group. More generally, we can consider cases in which $Q\not\simeq\mathbb{Z}_2^n$ and some of the order-two Verlinde lines correspond to $\vec\ell\not\in Q$ (e.g., see the ${SU}(2)$ at level one WZW model example in section \ref{defEx}). In this case we also  obtain the full Pauli group: $\vec\ell$ in \eqref{codeDZ} is any order-two element, and $\vec p$ is any representation in the Narain theory. Therefore, the charge-conjugation modular invariant knows about the full set of operations acting on the quantum code: {\it the genuine local operators correspond to stabilizers and the defect endpoint operators correspond to the errors.}

It is straightforward to extend this picture to the most general MPMIs when these CFTs admit a quantum code description. Clearly, to be an MPMI, we need every possible $\vec p$ and $\overline{\vec g+\vec p}$ to appear exactly once in \eqref{HZ}. Therefore, as we sum over $\vec g$ and take all $\vec p\in B_{\vec g}$, we produce all possible $\vec p\in K$. As a result, in the code we generate via \eqref{codeZ}, we get all possible powers of $Z$. The powers of $X$ are restricted since $\vec g\in Q$, and $Q$ is a proper subgroup of $K$.

However, the fields living at the end of the order-two Verlinde defects precisely make up the difference since \eqref{permD} now becomes
\begin{eqnarray}
Z_{\mathcal{T}_{\CM}}^{\vec\ell}(q,\bar q)&=&\sum_{\vec g\in H}\sum_{\vec p\in B_{\vec g}}\chi_{\vec p+\vec\ell}(q)\bar\chi_{\overline{\vec p+\vec g}}(q)\cr&=&\sum_{\vec g\in H}\sum_{\vec p\in B_{\vec g}}\chi_{\vec p+\vec\ell}(q)\bar\chi_{\overline{\vec g+\vec\ell+\vec p+\vec\ell}}(\bar q)~.
\end{eqnarray}
As a result, our CFT-code map in \eqref{codeZ} becomes
\be\label{codeDMZ}
\left\{\CO^{\vec\ell}_{\vec p+\vec \ell,\overline{\vec g+\vec p}}\right\}\, \leftrightarrow\,  X^{M (\vec \ell+\vec g)} \circ Z^{A (\vec p+\vec \ell) }~.
\ee
Since the fusion rules in \eqref{Vfusion} do not depend on the nature of $\CM$, we see that the number of order-two Verlinde defects is the same as in the charge-conjugation case. Therefore, upon including all order-two Verlinde lines, we get all possible Pauli group elements, and the corresponding errors that affect our stabilizer code.

Let us now consider the most general case \eqref{HZ}, which we can always write as in \eqref{MZ} with $\CT_M=\CT/Q$ (and discrete torsion $[\sigma]$). Note that in \eqref{MZ}, $\CM_{\vec p,\vec q}$ is a matrix with entries consisting of $0$'s and $1$'s (see Appendix B), and it will not generally be a permutation (i.e., the CFT will not be an MPMI).

As we will see in the next subsection, we have a smaller number of Verlinde lines when $\CT/Q$ is not an MPMI. However, we can still define enough order-two symmetries to recover the Pauli group from the corresponding defect fields (note that invertibility of the orbifolding procedure guarantees that, for each symmetry we gauge, there is a dual symmetry in the orbifolded theory).

To construct these extra symmetries, it suffices to associate signs with the primaries compatible with fusion (then all local correlation functions are invariant). In the Verlinde line case, we did this via \eqref{VL} and \eqref{ord2}. 

Since we have orbifolded in a way that respects $\CT$'s chiral algebra, $\CT/Q$ respects the fusion rules of $\CT$. More precisely, if we have operators in the orbifolded theory transforming in representations $(\vec p_1, \overline{\vec g_1 + \vec p_1})$ and $(\vec p_2, \overline{\vec g_2 + \vec p_2})$, then we  also have an operator transforming as $(\vec p_1+\vec p_2, \overline{\vec g_1+ \vec g_2 + \vec p_1+ \vec p_2})$. Technically, this statement follows from
\bea
S_{\vec h,\vec  p_1 + \vec p_2} ~ \Xi(\vec h,\vec g_1 + \vec g_2)&=&S_{\vec h,\vec  p_1} ~ \Xi(\vec h,\vec g_1)~ S_{\vec h,\vec p_2} ~ \Xi(\vec h,\vec g_2)\cr&=&1 ~, ~\quad\forall \vec h \in \DZ_2^k~,
\eea
where we have used the bicharacter property of both $S$ and $\Xi$ (see Appendix A). Therefore, $(\vec p, \overline{\vec g + \vec p})$ forms an abelian group under fusion (as it should since $\CT/Q$ is a Narain theory). Let us denote this group as $F$.

Now, after acting with some order-two symmetry, $\pi$ (i.e., inserting the corresponding topological defect, $D_{\pi}$, along a spatial cycle and computing the torus partition function), some of the $1$ entries in $\CM$ get flipped to $-1$ such that fusion is respected. Let us denote the matrix so obtained as $\CM_{\pi}$.

As in \eqref{modTransD}, to calculate the defect partition function, we have to perform an $S$ transformation to get $S^T\CM_{\pi}\bar S$. All the characters that we get from the defect partition functions for all possible order-two $\pi$ correspond to the non-zero entries of the matrix
\be\label{MSdef}
\sum_\pi S^T\CM_{\pi}\bar S= S^T \bigg (\sum_\pi \CM_\pi\bigg ) \bar S:=S^T\CM_{\Sigma}\bar S~,
\ee
where the sum is over all such symmetries, $\pi$.

Assigning signs to the primaries such that the fusion is respected is the same as choosing an irreducible representation of $F$ valued in $\pm 1$. The trivial representation acts trivially on the primaries. Therefore, for each $\pi$, we associate an irrep, ${\rm sign}\ \pi$. In order to find the non-zero entries of $\sum_\pi \CM_\pi$ we have to understand when
\be\label{sumS}
\sigma(x):=\sum_{\text{sign } \pi} \chi_{{\rm sign}\ \pi}(x)~,
\ee
is non-zero. Here, the sum is over the irreducible representations, ${\rm sign}\ \pi$, of $F$ valued in $\pm 1$, and $\chi_{{\rm sign}\ \pi}(x)$ is the character of ${\rm sign}\ \pi$ (not to be confused with the RCFT characters appearing in the partition function!) evaluated on a given element $x \in F$ (note that each element in $F$ represents a character combination $\chi_{\vec p}\bar\chi_{\overline{\vec g+\vec p}}\in Z_{\CT/Q,[\sigma]}$; we will denote this combination $(\vec p,\overline{\vec g+\vec p})$). 

To that end, suppose $F$ has a decomposition in terms of cyclic groups given by
\be
F \cong \DZ_{n_1} \otimes ... \otimes \DZ_{n_l}~.
\ee
Since we are treating CFT factors related to $A_{q^r}$ and $B_{q^r}$ as spectators, the $n_i$ are even.

We know that 
\be
\hat F= \hat \DZ_{n_1} \otimes ... \otimes \hat \DZ_{n_l} ~,
\ee
where $\hat F$ is the group of irreducible representations of $F$. In particular, the sign representations of $F$ are given by products of $\DZ_{n_i}$ sign representations. Choose a basis $\{e_1,\cdots,e_l\}$ for the cyclic groups; then, an element of $F$ is of the form $(e_1^{m_1},...,e_{l}^{m_l})$ for some integers $0\leq m_i \leq n_i-1$. Consider $\sigma(x)$ for some $x=(e_1^{m_1},...,e_{l}^{m_l}) \in F$. We know that ${\rm sign}\ \pi = {\rm sign}\ \pi_1 \otimes\cdots \otimes {\rm sign}\ \pi_{l}$, where ${\rm sign}\ \pi_i$ is a representation of $\DZ_{n_i}$ valued in $\pm 1$. Therefore
\bea
\sigma(x)=&&\sum_{{\rm sign}\ \pi_1,\cdots,\ {\rm sign}\ \pi_l} \chi_{\pi_1}(e_1^{m_1})\dots \chi_{\pi_l}(e_1^{m_l})=\cr && \prod_i \left(\sum_{{\rm sign}\ \pi_i} \left(\chi_{{\rm sign}\ \pi_i}(e_i)\right)^{m_1} \right)~.
\eea
Since the $n_i$ are all even and ${\rm sign}\ \pi_i$ is valued in $\pm 1$, we have $\chi_{{\rm sign}\ \pi_i}(e_i)=\pm 1 \forall i$. Therefore, we find
\be
\sigma(x)= \prod_i (1^{m_i}+(-1)^{m_i})= \begin{cases}
2^l~, &\text{iff}\ m_i\in2\mathbb{Z}\ \forall i\\
0~, &\text{otherwise}~.
\end{cases}
\ee
Now, suppose $x=(e_1^{m_1},...,e_{l}^{m_l})$ is an element of the group $F$ such that all $m_i$ are even. Then there exists some other element $y \in F$ such that $y^2=x$. Recall that an element of $F$ represents a character combination in the partition function denoted by $(\vec p, \overline{\vec g + \vec p})$.
Adding this element to itself gives $(2 \vec p, 2 \overline{\vec p} )$ (since $\vec g$ is order two). Therefore, if $x \in F$ has only even $m_i$, $x=(2 \vec p, 2 \overline{\vec p})$.

As a result, the matrix $\CM_{\Sigma}$ defined in \eqref{MSdef} is a matrix with entries valued in $\{0,2^l\}$, where the only non-zero entries correspond to $(2 \vec p, 2 \overline{\vec p})$. In other words
\bea\label{sumZ}
\sum_{\pi} Z_{\CT/Q,[\sigma]}(D_{\pi})&=&\sum_{\pi}\CM_{\pi;\vec p,\vec q}\chi_{\vec p}(q)\bar\chi_{\vec q}(\bar{\vec q})\cr&=& 2^l \sum_{\vec 2 p} \chi_{2 \vec p}(q) \bar \chi_{\overline{2 \vec p}}(\bar q)~.
\eea
Note that the case $\pi=1$ gives the partition function without a defect. As a result, $\chi_{2 \vec p} \bar \chi_{\overline{2 \vec p}}$ is a term in this partition function, and we know that $2 \vec p$ has to satisfy \eqref{pconst2} for $\vec g=0$. That is, the CS Wilson line corresponding to $2 \vec p$ should braid trivially with all $\vec h \in \DZ_2^k$. 

We want to show that the sum of defect partition functions $\sum_{\pi} Z_{\CT/Q,[\sigma]}^{\pi}$ (coming from applying a modular transformation to \eqref{sumZ}) contains all possible characters of the form $\chi_{\vec p} \bar \chi_{\overline{\vec g + \vec p}}$, where $\vec g$ is order two, so that we get the full Pauli group from it. To that end, consider
\bea\label{StransSumZ}
\sum_{\pi} Z_{\CT/Q,[\sigma]}^{\pi}&=& 2^l \sum_{2\vec p} \sum_{\vec i,\vec j} S_{2 \vec p, \vec i} \bar S_{2\bar{\vec p},\vec j} \chi_{\vec i} \bar \chi_{\overline{\vec j}}\cr&=& 2^l \sum_{2\vec p} \sum_{\vec i,\vec j} S_{2 \vec p, (\vec i-\vec j)} \chi_{\vec i} \bar \chi_{\overline{\vec j}}\cr&=&2^l \sum_{2\vec p} \sum_{\vec i,\vec j} S_{\vec p, 2(\vec i-\vec j)} \chi_{\vec i} \bar \chi_{\overline{\vec j}}~.
\eea
It is clear that if $(\vec i-\vec j)$ is order two, then $S_{\vec p, 2(\vec i-\vec j)}=1 \ \forall \vec p$. Therefore, the character $\chi_{\vec i} \bar\chi_{\bar{\vec j}}$ contributes non-trivally to the sum for any $\vec i, \vec j$ satisfying the constraint that $\vec i-\vec j$ is order two. These characters correspond to
\be 
\label{gendefPauli}
X^{M({\vec i - \vec j})} \circ Z^{A\vec i}~.
\ee 
Since $\vec i-\vec j$ is any order-two element, and $\vec i$ is arbitrary (though choosing $\vec i$ fixes $\vec j$ mod 2), we find that these defect fields give the full Pauli group. {\it This ends our proof and shows that all code CFTs contain all possible errors via order-two defects.}

\subsection{Verlinde Subgroup of the Pauli Group}\label{Verlinde}
In this section, we define a \lq\lq Verlinde subgroup" of $\mathcal{P}_n$. This subgroup can be constructed from any code RCFT. It is defined as follows.

\noindent
{\bf Definition:} The Verlinde subgroup, $\mathcal{V}_{\CT/Q}$, is the subgroup of $\mathcal{P}_{\CT/Q}$ coming from all stabilizers that are related to (1) CFT  local fields and (2) fields living at the end of order-two Verlinde lines.

\noindent
Note that, by construction $\CS_{\CT/Q}\subseteq\mathcal{V}_{\CT/Q}\subseteq\mathcal{P}_{\CT/Q}$. Physically, the ratio
\begin{equation}
r_{\CT/Q}:=2^{-n}{|\mathcal{P}_{\CT/Q}|\over|\mathcal{V}_{\CT/Q}|}~, \ \ \ 2^{-n}\le r\le 1~,
\end{equation}
measures how well the continuous symmetries of the Narain CFT corresponding to an $n$-qubit code are able to detect an error. For example, in the charge conjugation modular invariant or any of the MPMIs, $r_{\CT/Q}=2^{-n}$, which is the smallest value possible. This is because the Verlinde subgroup corresponds to the full Pauli group. Any Verlinde line, $\CL_{\vec\ell}$, commutes with the chiral algebra, since $\bar S_{\vec \ell\vec 0}/\bar S_{\vec 0\vec 0}=1$ in \eqref{VL}, and so the corresponding continuous symmetry currents are acted upon trivially by the Verlinde lines. In this sense, the continuous symmetry currents cannot detect errors associated with these defects.

What about more general theories? These theories are not MPMIs. However, it turns out that, if we enlarge the chiral algebras as much as possible, any orbifold theory we can construct using our methods above is a permutation modular invariant with respect to this larger algebra (see Appendix D). We can then define a Verlinde subgroup for any of our orbifold theories. Moreover, as we show in Appendix D, if we enlarge the chiral algebra, then, $r_{\CT/Q}>2^{-n}$, and the error detection ability of the continuous symmetry currents improves. In the most extreme cases, we get CFTs that are products of left moving meromorphic and right moving anti-meromorphic CFTs. These types of theories have  $r_{\CT/Q}=1$, and their continuous symmetries are able to fully detect errors.

\subsection{Examples}\label{defEx}

\subsubsection{Pauli group from $R=1$ compact boson}

The $R=1$ compact boson has a charge conjugation partition function which is an MPMI. Therefore, our general discussion on Pauli groups from MPMIs can be readily applied to this case. To that end, consider
\be
Z_{\CT}=\chi_0\bar\chi_0+\chi_{2}\bar\chi_{2}+\chi_1\bar\chi_3+\chi_{3}\bar\chi_{ 1}~.
\ee
Recall that the bulk operators are mapped to the 1-qubit stabilizer code, gen$(Z)$. This CFT has a $\DZ_2$ symmetry generated by the Verlinde line, $\CL_2$. Inserting this line in the partition function, we can calculate the defect partition function using \eqref{chargedef}
\be
Z^{\vec\ell=2}_{\CT}= \chi_0\bar\chi_2+\chi_{2}\bar\chi_{0}+\chi_1\bar\chi_1+\chi_{3}\bar\chi_{3}~.
\ee
Using \eqref{codeDZ}, the defect operators are mapped to Pauli group elements as follows
\be
X \leftrightarrow \{\CO_{(0,2)}^{\vec\ell=2}\}, \{\CO_{(2,0)}^{\vec\ell=2}\}~, \ Y \leftrightarrow \{\CO_{(1,1)}^{\vec\ell=2}\}, \{\CO_{(3,3)}^{\vec\ell=2}\} ~.
\ee
Therefore, the bulk operators along with the defect operators give us the full Pauli group, $\mathcal{P}_\CT$. Since the $X$ and $Y$ Pauli matrices correspond to defect operators living at the end of an order-two Verlinde line, the Verlinde subgroup, $\mathcal{V}_{\CT}$, is the full Pauli group. 

\subsubsection{Pauli group from $R=\sqrt{\frac{2}{2^{s-1}}}$ compact boson}

Recall that the $R=\sqrt{\frac{2}{2^{s-1}}}$ compact boson has the charge conjugation partition function
\be
Z_{\CT}= \sum_{p \in \DZ_{2^{s}}} \chi_{p} \bar \chi_{\bar p}~, 
\ee
We know that the CFT local operators are mapped to the qubit stabilier code generated by $Z$. This CFT has a $\DZ_2$ symmetry generated by the Verlinde line, $\CL_{2^{s-1}}$. Inserting this line in the partition function, we can calculate the defect partition function using \eqref{chargedef}
\be
Z^{\vec\ell=2^{s-1}}_{\CT}= \sum_{p \in \DZ_{2^{s}}} \chi_{p+2^{s-1}} \bar \chi_{\bar p}~, 
\ee
Using \eqref{codeDZ}, the defect operators are mapped to Pauli group elements as follows
\bea
X \leftrightarrow \{\CO_{p+2^{s-1},\bar p}^{\vec\ell=2^{s-1}}\} ~,~ p=\text{0 \text{ mod } 2}~, \cr
Y \leftrightarrow \{\CO_{p+2^{s-1},\bar p}^{\vec\ell=2^{s-1}}\} ~,~ p=\text{1 \text{ mod } 2} ~.
\eea
Therefore, the local operators along with the defect operators at the end of the order-two Verlinde line $\CL_{2^s}$ gives us the full Pauli group. 

Now let us consider the CFT with partition function 
\be
\label{RcbZ2part}
Z_{\CT/\DZ_2}= \sum_{p= 0\text{ mod }2, p \in \DZ_{2^{s}}} \chi_{p} \bar \chi_{\bar p} + \chi_{p} \bar \chi_{\overline{2^s + p}}~, 
\ee
obtained from the $R=2^{2-s\over2}$ CFT by orbifolding the $\DZ_2$ symmetry generated by $\CL_{2^{s-1}}$. Recall that the genuine local operators in this CFT are mapped to the stabilizer code generated by $X$ (for $s>2$). 

This CFT has a $\DZ_2$ symmetry generated by a line defect, say $D_\pi$, which acts on the primary operators as follows
\be
\{\CO_{v,\bar v}\} \rightarrow \{\CO_{v,\bar v}\}, ~ \{\CO_{v,\overline{2^{s-1}+v}}\} \rightarrow - \{\CO_{v,\overline{2^{s-1}+v}}\}
\ee
Using a modular $S$ transformation, we can find the defect partition function
\be
Z_{\CT/\DZ_2}(D_\pi)= \sum_{p= 1\text{ mod }2, p \in \DZ_{2^{s}}} \chi_{p} \bar \chi_{\bar p} + \chi_{p} \bar \chi_{\overline{2^{s-1} + p}}~, 
\ee
Using \eqref{codeDZ}, the defect operators are mapped to Pauli group elements as follows
\bea
Z \leftrightarrow \{\CO_{p,\bar p}^{D_\pi}\},~ Y \leftrightarrow \{\CO_{p+2^{s-1},\bar p}^{D_\pi}\} ~,
\eea
where $p=1 \text{ mod }2$.
Therefore, we find that the local operators of the CFT along with the defect operators give us the full Pauli group.

Note that the partition function \eqref{RcbZ2part} is clearly not an MPMI. In this case we get the non-trivial group $E=\{0,2^{s-1}\}$ defined in section \ref{Verlinde}. Therefore, using \eqref{charactermap}, we can enlarge the chiral algebra as follows.
\be
\tilde \chi_{0}= \chi_{0}+ \chi_{2^{s-1}},~ \tilde \chi_{\rho}= \chi_{\rho}+ \chi_{\rho+2^{s-1}}
\ee
where $\rho$ is a a representative of the orbit $\{v,v+2^{s-1}\}, v =0 \text{ mod }2, v\in \DZ_{2^{s}}$.
With respect to this enlarged chiral algebra, we have the partition function
\be
Z_{\CT/\mathbb{Z}_2}= \sum_{\rho} \tilde \chi_{\rho} \bar{\tilde{\chi}}_{\bar \rho}~.
\ee
Therefore, we have Verlinde lines labelled by the primaries $\rho$. However, we don't have any non-trivial order-two Verlinde lines. Therefore, the Verlinde subgroup is same as the stabilizer group.    

\subsubsection{Pauli group from $\widehat{Spin(16)}_1$ CFT}

Recall that the $\widehat{Spin(16)}_1$ CFT has the charge-conjugation partition function
\be
Z_{\CT}=\chi_{(0,0)} \bar \chi_{(0,0)}+\chi_{(0,1)} \bar \chi_{(0,1)}+\chi_{(1,0)} \bar \chi_{(1,0)}+\chi_{(1,1)} \bar \chi_{(1,1)}~,
\ee
and the bulk operators are mapped to the 2-qubit stabilizer code gen$(I \otimes Z, Z \otimes I)$. This CFT has $\DZ_2 \times \DZ_2$ 0-form symmetry generated by the Verlinde lines $\CL_{(0,1)}$ and $\CL_{(1,0)}$.  Inserting these lines in the parition function, we obtain the following defect partition functions via \eqref{chargedef}
\bea
Z_{\CT}^{(0,1)}=\chi_{(0,1)} \bar \chi_{(0,0)}&+&\chi_{(0,0)} \bar \chi_{(0,1)}+\chi_{(1,1)} \bar \chi_{(1,0)}\cr
&&\ +\chi_{(1,0)} \bar\chi_{(1,1)}~, \cr
Z_{\CT}^{(1,0)}=\chi_{(1,0)} \bar \chi_{(0,0)}&+&\chi_{(1,1)} \bar\chi_{(0,1)}+\chi_{(0,0)} \bar\chi_{(1,0)} \cr
&&\ +\chi_{(0,1)} \bar\chi_{(1,1)}~, \cr
Z_{\CT}^{(1,1)}=\chi_{(1,1)} \bar\chi_{(0,0)}&+&\chi_{(1,0)} \bar\chi_{(0,1)}+\chi_{(0,1)} \bar\chi_{(1,0)} \cr
&&\ +\chi_{(0,0)} \bar\chi_{(1,1)}~.
\eea
Using \eqref{codeDZ}, the defect operators are, in turn, mapped to Pauli group elements
\be
 Z\otimes X, I\otimes X, Z\otimes Y, I\otimes Y ~,
\ee
\be
 X\otimes Z, Y\otimes Z, X\otimes I, Y\otimes I ~,
\ee
\be
 Y\otimes Y, X\otimes Y, Y\otimes X, X\otimes X ~ .
\ee
Therefore, the bulk operators along with the defect operators give us the full Pauli group $\mathcal{P}_\CT$. Since all defect operators live at the end of order-two Verlinde lines, the Verlinde subgroup, $\mathcal{V}_{\CT}$, is the full Pauli group. 

Now let us consider the CFT with partition function 
\bea
\label{partfunlab}
Z_{\CT/Q_1}= \chi_{(0,0)}\bar\chi_{(0,0)} &+& \chi_{(0,0)}\bar\chi_{(0,1)} + \chi_{(0,1)}\bar\chi_{(0,0)}\cr &&\ + \chi_{(0,1)}\bar\chi_{(0,1)} ~,
\eea
obtained from the $\widehat{Spin(16)}_1$ CFT by orbifolding the $Q_1$ symmetry generated by $\CL_{(0,1)}$. Recall that the bulk operators are mapped to the 2-qubit stabilizer code gen$(Z \otimes I, I \otimes X)$. This CFT has order-two symmetries generated by $D_{\pi_1}$ and $D_{\pi_2}$. $D_{\pi_1}$ acts on the primaries as
\bea
\{\CO_{(0,0),(0,1)}\} &\rightarrow& -\{\CO_{(0,0),(0,1)}\}~, \cr \{\CO_{(0,1),(0,0)}\} &\rightarrow& -\{\CO_{(0,1),(0,0)}\}~,
\eea
and trivially on $\{\CO_{(0,0),(0,0)}\}$ and $\{\CO_{(0,1),(0,1)}\}$. $D_{\pi_2}$ acts on the primaries as
\bea
\{\CO_{(0,1),(0,0)}\} &\rightarrow & -\{\CO_{(0,1),(0,0)}\} \text{ and } \cr \{\CO_{(0,1),(0,1)}\} &\rightarrow & -\{\CO_{(0,1),(0,1)}\}~,
\eea
and trivially on $\{\CO_{(0,0),(0,0)}\}$ and $\{\CO_{(0,0),(0,1)}\}$.   

Using a modular $S$ transformation, we can find the defect partition functions
\bea 
Z_{\CT/Q_1}(D_{\pi_1})= \chi_{(1,0)}\bar\chi_{(1,0)} &+& \chi_{(1,1)}\bar\chi_{(1,1)} + \chi_{(1,1)}\bar\chi_{(1,0)}\cr &&\ + \chi_{(1,0)}\bar\chi_{(1,1)} ~, \cr
Z_{\CT/Q_2}(D_{\pi_2})= \chi_{(1,0)}\bar\chi_{(0,0)} &+& \chi_{(1,0)}\bar\chi_{(0,1)} + \chi_{(1,1)}\bar\chi_{(0,0)}\cr &&\ + \chi_{(1,1)}\bar\chi_{(0,1)} ~, \cr
Z_{\CT/Q_3}(D_{\pi_1\pi_2})= \chi_{(0,0)}\bar\chi_{(1,0)} &+& \chi_{(0,0)}\bar\chi_{(1,1)} + \chi_{(0,1)}\bar\chi_{(1,0)}\cr &&\ + \chi_{(0,1)}\bar\chi_{(1,1)} ~. 
\eea
Using \eqref{gendefPauli}, the defect operators are, in turn, mapped to Pauli group elements  
\be
I \otimes Z, Z \otimes Z, X \otimes Y, I \otimes Y ~,
\ee
\be
X \otimes Z, X \otimes Y, Y \otimes Y, Y \otimes Z ~,
\ee
\be
X \otimes I, X \otimes X, Y \otimes X, Y \otimes I ~.
\ee
Therefore, the bulk fields along with the defect fields give us the full 2-qubit Pauli group. 

The Verlinde subgroup in this case is the same as the stabilizer group. To understand this statement, note that the partition function \eqref{partfunlab} is clearly not an MPMI. In this case we get the non-trivial group $E=\{(0,0),(0,1)\}$ defined in section \ref{Verlinde}. Therefore, using \eqref{charactermap}, we can enlarge the chiral algebra as follows.
\be\label{LargerChi}
\tilde \chi_{\vec 0}= \chi_{(0,0)}+ \chi_{(0,1)}~. 
\ee
With respect to this enlarged chiral algebra, we have
\be
Z_{\CT/Q_1}= \tilde \chi_{\vec 0} \bar \tilde \chi_{\vec 0}~.
\ee
We get a meromorphic RCFT times an anti-meromorphic RCFT. In this case we don't have any non-trivial Verlinde lines, and $\mathcal{V}_{\CT/Q_1}=\mathcal{S}_2$.

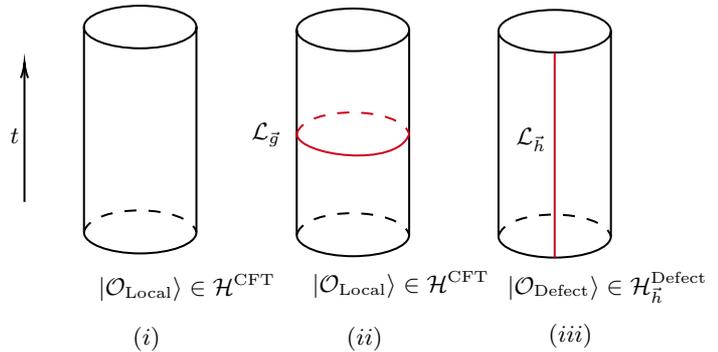
\begin{figure}
\centering

\tikzset{every picture/.style={line width=0.75pt}} 

\begin{tikzpicture}[x=0.75pt,y=0.75pt,yscale=-0.7,xscale=0.55]

\draw   (187.55,37.3) -- (187.55,184.7) .. controls (187.55,193.15) and (164.71,200) .. (136.55,200) .. controls (108.38,200) and (85.55,193.15) .. (85.55,184.7) -- (85.55,37.3) .. controls (85.55,28.85) and (108.38,22) .. (136.55,22) .. controls (164.71,22) and (187.55,28.85) .. (187.55,37.3) .. controls (187.55,45.75) and (164.71,52.6) .. (136.55,52.6) .. controls (108.38,52.6) and (85.55,45.75) .. (85.55,37.3) ;
\draw  [dash pattern={on 4.5pt off 4.5pt}]  (85.55,184.7) .. controls (92.55,163.4) and (180.55,164.4) .. (187.55,184.7) ;
\draw   (380.55,39.3) -- (380.55,186.7) .. controls (380.55,195.15) and (357.71,202) .. (329.55,202) .. controls (301.38,202) and (278.55,195.15) .. (278.55,186.7) -- (278.55,39.3) .. controls (278.55,30.85) and (301.38,24) .. (329.55,24) .. controls (357.71,24) and (380.55,30.85) .. (380.55,39.3) .. controls (380.55,47.75) and (357.71,54.6) .. (329.55,54.6) .. controls (301.38,54.6) and (278.55,47.75) .. (278.55,39.3) ;
\draw [color={rgb, 255:red, 208; green, 2; blue, 27 }  ,draw opacity=1 ]   (278.55,114.3) .. controls (280.55,130) and (369.55,139) .. (380.55,114.3) ;
\draw [color={rgb, 255:red, 208; green, 2; blue, 27 }  ,draw opacity=1 ] [dash pattern={on 4.5pt off 4.5pt}]  (278.55,114.3) .. controls (285.55,93) and (373.55,94) .. (380.55,114.3) ;
\draw  [dash pattern={on 4.5pt off 4.5pt}]  (278.55,186.7) .. controls (285.55,165.4) and (373.55,166.4) .. (380.55,186.7) ;
\draw   (563.55,40.3) -- (563.55,187.7) .. controls (563.55,196.15) and (540.71,203) .. (512.55,203) .. controls (484.38,203) and (461.55,196.15) .. (461.55,187.7) -- (461.55,40.3) .. controls (461.55,31.85) and (484.38,25) .. (512.55,25) .. controls (540.71,25) and (563.55,31.85) .. (563.55,40.3) .. controls (563.55,48.75) and (540.71,55.6) .. (512.55,55.6) .. controls (484.38,55.6) and (461.55,48.75) .. (461.55,40.3) ;
\draw  [dash pattern={on 4.5pt off 4.5pt}]  (461.55,187.7) .. controls (468.55,166.4) and (556.55,167.4) .. (563.55,187.7) ;
\draw [color={rgb, 255:red, 208; green, 2; blue, 27 }  ,draw opacity=1 ]   (512.55,55.6) -- (512.55,203) ;
\draw    (31.55,163) -- (31.55,65) ;
\draw [shift={(31.55,63)}, rotate = 90] [color={rgb, 255:red, 0; green, 0; blue, 0 }  ][line width=0.75]    (10.93,-3.29) .. controls (6.95,-1.4) and (3.31,-0.3) .. (0,0) .. controls (3.31,0.3) and (6.95,1.4) .. (10.93,3.29)   ;

\draw (96,212.4) node [anchor=north west][inner sep=0.75pt]    {$\ket{\mathcal{O}_{\text{Local}}}\in \mathcal{H}^{\text{CFT}}$};
\draw (235,104.4) node [anchor=north west][inner sep=0.75pt]    {$\CL_{\vec g}$};
\draw (290,210.4) node [anchor=north west][inner sep=0.75pt]    {$\ket{\mathcal{O}_{\text{Local}}} \in \mathcal{H}^{\text{CFT}}$};
\draw (467,211.4) node [anchor=north west][inner sep=0.75pt]    {$\ket{\mathcal{O}_{\text{Defect}}}\in \mathcal{H}_{\vec h}^{\text{Defect}}$};
\draw (16,108.4) node [anchor=north west][inner sep=0.75pt]    {$t$};
\draw (127,250.4) node [anchor=north west][inner sep=0.75pt]    {$( i)$};
\draw (321,250.4) node [anchor=north west][inner sep=0.75pt]    {$( ii)$};
\draw (504,248.4) node [anchor=north west][inner sep=0.75pt]    {$( iii)$};
\draw (475,107.4) node [anchor=north west][inner sep=0.75pt]    {$\CL_{\vec h}$};

\end{tikzpicture}

\caption{The CFT on $S^1\times\mathbb{R}$: (i) The code subspace maps to the CFT states corresponding to genuine local operators (ii) A CFT logical operation: wrapping the spatial slice with a symmetry defect, $\CL_{\vec g}$, implements the symmetry on $\CH_{\text{Bulk}}^{\text{CFT}}$ (at the level of the code, the logical operation is trivial). (iii) The complement of the code subspace in the $n$-qubit Hilbert space: a state in the $\CL_{\vec h}$-defect Hilbert space (here $2\vec h=\vec 0$).}
\end{figure}

\section{The qubit Hilbert space /  CFT Hilbert space map}
We have constructed a map that relates the stabilizers and error operations acting on $n$ qubits to an infinite number of genuine local and defect endpoint operators in very general Narain RCFTs. How then should we map the $n$-qubit Hilbert space, $\CH_n$, to the infinite-dimensional CFT Hilbert space? 

Let us first consider the code subspace, $\CC_n \subset \CH_n$. It is defined as the space invariant under the action of the stabilizer group. In our case it is one dimensional. To find the corresponding CFT states, we look for the space which is closed under action of genuine local CFT operators, since these operators correspond to stabilizers under the map $\mu$ \eqref{MuMap2}. By the state-operator correspondence, this is nothing but the CFT Hilbert space
\begin{equation}
\label{codes}
\mu(\CH^{\rm CFT})=\CC_n~.
\end{equation}
Note that, at the level of the CFT Hilbert space, logical operations are non-trivial, but they become trivial after the action of $\mu$ \eqref{codes}.

Next, what are the $2^n-1$ states in the complement of $\CC_n$ inside the $n$-qubit Hilbert space on the CFT side? The natural choice is that these correspond to the $2^n-1$ different defect Hilbert spaces, $\CH^{\rm Defect}_i$, associated with the defect endpoint fields we interpreted as errors in section \ref{defects},
\begin{equation}
\label{complement}
\mu(\CH^{\rm Defect}_i)=\CC^c_n := \CH_n\backslash  \, \CC_n~. 
\end{equation}
The basic property of $\CC^c_n$ is that error operations acting on $\CC_n$ produce states in the complement.  This property is respected by $\mu$: inserting a defect endpoint operator takes us from the bulk CFT Hilbert space to the corresponding defect Hilbert space. We illustrate our proposal \eqref{codes} and \eqref{complement} in Figs. 3 (i)-(iii).

\section{Discussion and Conclusions}
We have proposed a map from very general rational Narain CFTs (including defects), and their associated CS theories, to stabilizer codes. This construction includes the theories discussed in \cite{dymarsky2021quantum,dymarsky2021quantum2} as a special case, and provides a CFT picture of the code space states and errors reminiscent of the toric code construction \cite{KitaevToric}. 

Our CFT to stabilizer map works as follows. First, we pick a Narain theory with a particular chiral algebra and construct the charge conjugation modular invariant. We then consider all orbifolds by $Q=\mathbb{Z}_2^k$ subgroups of the 0-form flavor symmetry that come from 3d CS 1-form symmetries with vanishing 't Hooft anomalies (this condition ensures the stabilizer group is abelian \eqref{stabCond}) and relate genuine local operators to stabilizer generators \eqref{codeZ}. Under this map, operators sitting at the ends of line defects are mapped to Pauli operators acting on logical qubits. Accordingly, the whole bulk CFT Hilbert space is mapped to the code subspace \eqref{codes}, while defect Hilbert spaces are mapped to the complement of the code subspace in the $n$-qubit Hilbert space \eqref{complement}.

Note that, while the map  is unambiguous, it can lead to the same CFT having different codes associated with it because certain CFTs can be considered rational with respect to multiple chiral algebras. For example, the $\widehat{Spin(16)}_1/\mathbb{Z}_2$ orbifolds discussed in section \ref{Examples1} can be interpreted as corresponding to two different chiral algebras. If we run our map with the smaller  chiral algebra $V_{\rm min}=V_{\widehat{Spin(16)}_1}$, we produce the sequence of RCFT / code relations discussed in the text. On the other hand, if we use  maximal  chiral algebra, $V_{\rm max}$, described around \eqref{LargerChi}, then the $\widehat{Spin(16)}_1/\mathbb{Z}_2$ orbifolds correspond to trivial 0-qubit codes, as follows from triviality of ${\rm Rep}(V_{\rm max})$, see the discussion below \eqref{LargerChi}.

Within our construction, it is natural  to ask if we can construct a CFT  starting from a given stabilizer code. Since there might be different CFTs related to that code, it is clear that we need extra data. Starting from the stabilizers, we can choose a group $Q$, and a 2-cocycle, $\sigma\in H^2(Q,U(1))$, compatible with the code. To reconstruct the CFT requires choosing a chiral algebra such that the charge conjugation modular invariant with that chiral algebra admits a non-anomalous 0-form symmetry isomorphic to $Q$. Taking the $Q$-orbifold of this CFT with discrete torsion, $\sigma$, gives  a CFT corresponding to the quantum code in question. An alternative approach  is to define a Narain lattice  starting from a quantum code. One particular receipe is given by the ``new Construction A" of \cite{dymarsky2021quantum}, which  can be used to construct  orbifolds of the charge conjugation modular invariant with Rep$(V)=A_{4}^{n_{A_4}}$ for arbitrary integer $n_{A_4}$. There are, of course, other constructions leading to other CFTs for the same or other codes. For example, the Narain lattice \eqref{latticesu2} for the $SU(2)$ WZW model at level one can be generalized to yield CFTs with Rep$(V)=A_2^{n_{A_2}}$ for arbitrary integer $n_{A_2}>0$.   

Our work opens a number of new directions to explore:
\begin{itemize}
\item{We have emphasized that different CFTs can be associated with the same code. It is natural to ask if the space of CFTs related to a particular code admits additional structure.  One possible idea is to relate these theories by RG flow, or perhaps, some other form of coarse-graining. More broadly, these  theories may comprise  deformation classes reminiscent of topological modular forms in 2d $\CN=(0,1)$ theories, see e.g.~\cite{Tachikawa:2021mvw,ST1,ST2}. 

An alternative idea comes from  the example discussed below \eqref{MuMap}, where different CFTs mapping to the same code correspond to CS theories that are related by Galois conjugation \cite{Buican:2021axn,Harvey:2019qzs}. A natural question to ask is if more general Galois transformations always relate theories corresponding to the same code. 

Finally, when a $d$-dimensional QFT is invariant under gauging a $(d-2)/2$-form symmetry, one finds a non-invertible \lq\lq duality" defect \cite{Choi:2021kmx,Kaidi:2021xfk}. In 2d, these defects arise when a theory is invariant under gauging a zero-form symmetry, as in the case of the $R=1$ compact boson (see also \cite{TYW}). In this theory, we saw that the codes before and after gauging the $Q=\mathbb{Z}_2$ symmetry are equivalent. The codes before and after gauging are also equivalent for $R=\sqrt{2\over k}$ (for $k>2$) even though the theories are not. This result begs the question of whether code equivalences correspond, in the absence of an equivalence under gauging, to the existence of more general defects.
}

\item{The construction of this paper can be extended in many possible ways. 
In the discussion below \eqref{linmap}, the factors of  $A_{q^r}$ and $B_{q^r}$ in \eqref{pval}
are mapped into trivial (zero qubit) codes.  Quite naturally, these factors can be associated with qudit codes with $d=q$, where $d=2$ is the qubit case \cite{toappear}.
Another possible generalization comes from the choice of orbifold group, $Q$, in \eqref{Horbgroup} and, implicitly, a choice of stabilizer in \eqref{codeZ} for RCFTs corresponding to CS theories with $E_{2^r}$ and $F_{2^r}$ factors. Yet another natural generalization would be to include theories with non-abelian fusion rules. In this way, one may hope to extend our construction to all RCFTs. Going in a different direction, general CFT relations to codes are likely to extend beyond RCFTs to include non-rational ``finite'' theories \cite{Dymarsky:2021xfc}.

The broad program we are advocating here is to identify a generalization of codes which can be associated with general 2d CFTs.}

\item Relations to codes provide a powerful way to write CFT torus partition functions in terms of code enumerator polynomials. This relation applies to all CFTs discussed in this paper and can be extended to higher-genus partition functions \cite{Henriksson:2021qkt}. In this way, modular bootstrap constraints can be reformulated in terms of much simpler algebraic properties of enumerator polynomials, leading to a new approach to the modular bootstrap \cite{dymarsky2021quantum2}. Our work emphasized the importance of defects in the context of codes. We therefore surmise that codes will prove  useful as a new tool for the program of bootstrapping CFTs with defects (e.g., see \cite{TYW}). Since defects are also closely related to boundaries, we expect codes to have direct implications for bootstrapping in the presence of boundaries \cite{Collier:2021ngi}. Intriguingly, conformal boundaries are also related to gapped boundaries of the bulk TQFT \cite{cardy}. Therefore, it will be interesting to explore the role of quantum codes in describing and classifying gapped boundaries as in \cite{Kaidi:2021gbs}.

\item The physical meaning of quantum codes outlined in our paper, namely that the code subspace is related to the Hilbert space of CFT local operators, while errors correspond to defect endpoint operators,  has a natural holographic interpretation. Our theories are dual to 3d CS, where the code subspace and errors have a clear geometric meaning. We raise the question of making an explicit connection with the quantum codes, which define the space of low-energy bulk states in the context of holographic quantum gravity \cite{QECCAdSCFT}.

\end{itemize}

\acknowledgements{We thank A.~Gerasimov, I.~Runkel, A.~Shapere, and S.~Wood for discussions and correspondence. M.B. and R.R. are grateful to the Galileo Galilei Institute for a stimulating environment during the workshop \lq\lq Topological Properties of Gauge Theories and Their Applications to High-Energy and Condensed Matter Physics," where part of this work was completed. R.R. thanks ICMS for hospitality during the \lq\lq GCS: Kick-off Meeting." M.B.'s work was funded by the Royal Society under the grant \lq\lq Relations, Transformations, and Emergence in Quantum Field Theory." M.B.'s and R.R.'s work was funded by the Royal Society under the grant \lq\lq New Aspects of Conformal and Topological Field Theories Across Dimensions" and the STFC under the grant \lq\lq String Theory, Gauge Theory and Duality." A.D. is supported by the NSF under grant PHY-2013812.}

\appendix 

\subsection*{Appendix A: $S$ and $\Xi$ are bicharacters}\label{AppA}

In this appendix, we will show that both $S$ and $\Xi$ are bicharacters. To prove this, we need the following equations satisfied by $F(\vec p, \vec q, \vec r)$ and $R(\vec p, \vec q)$. 
\bea
\label{hexagons}
\frac{F(\vec q, \vec p, \vec r)}{F(\vec p, \vec q, \vec r)F(\vec q, \vec r, \vec p)}= \frac{R(\vec p, \vec q+ \vec r)}{R(\vec p, \vec q)R(\vec p, \vec r)}~,\cr
\frac{F(\vec p, \vec q, \vec r)F(\vec r, \vec p, \vec q)}{F(\vec p, \vec r, \vec q)}= \frac{R(\vec p+ \vec q, \vec r)}{R(\vec p, \vec r)R(\vec q, \vec r)}~.
\eea
These are known as the Hexagon equations \cite{mooreseiberg}. The modular $S$ matrix can be written in terms of $R$ as
\be
S_{\vec p, \vec q}= R(\vec p, \vec q)R(\vec q, \vec p)~.
\ee
We have
\bea
S_{\vec p, \vec q}S_{\vec p, \vec r}&=& R(\vec p, \vec q)R(\vec q, \vec p) R(\vec p, \vec r)R(\vec r, \vec p)\cr
&=& R(\vec p, \vec q+ \vec r)R(\vec q + \vec r, \vec p)= S_{\vec p, \vec q + \vec r}~,
\eea
where in the second equality we used \eqref{hexagons}. A similar argument can be used to show that $S_{\vec p, \vec r}S_{\vec q, \vec r}=S_{\vec p + \vec q, \vec r}$. This shows that the modular $S$ matrix is a bicharacter.

Consider the expression for $\Xi$ in terms of $R$, the 2-cochain $\tau$  and the 2-cocycle $\sigma$.
\be
\Xi(\vec g, \vec h)= R(\vec g, \vec h) \frac{\tau(\vec g, \vec h)\sigma(\vec g, \vec h)}{\tau(\vec h, \vec g)\sigma(\vec h, \vec g)}~.
\ee 
Recall that $\Xi$ is defined on a subgroup $Q$ of $K$ on which $F$ is trivial in cohomology. In fact, we can choose a gauge in which $F(\vec g, \vec h, \vec k)=1 ~ \forall \vec g , \vec h, \vec k \in Q$. Then $\tau(\vec g, \vec h)$ can be set to $1$ for all $\vec g, \vec h \in Q$. Therefore, we have
\bea
\Xi(\vec g, \vec h)\Xi(\vec g, \vec k)= R(\vec g, \vec h) R(\vec g, \vec k) \frac{\sigma(\vec g, \vec h)}{\sigma(\vec h, \vec g)} \frac{\sigma(\vec g, \vec k)}{\sigma(\vec k, \vec g)}\cr
= R(\vec g, \vec h + \vec k) \frac{\sigma(\vec g, \vec h+ \vec k)}{\sigma(\vec h+ \vec k, \vec g)}= \Xi(\vec g, \vec h+ \vec k)~,
\eea
where in the second equality above we used the property that for any 2-cocycle $\sigma$, $\frac{\sigma(\vec g, \vec h)}{\sigma(\vec h, \vec g)}$ is a bicharacter. A similar argument can be used to show that $\Xi(\vec g,\vec k)\Xi(\vec h, \vec k)= \Xi(\vec g+\vec h, \vec k)$. This shows that $\Xi$ is a bicharacter. 

\subsection*{Appendix B: Properties of $Z_{\CT/Q,[\sigma]}$} \label{AppB}

Let us discuss some properties of $Z_{\CT/Q,[\sigma]}$ which will be useful for our arguments. To that end, consider the general expression for $Z_{\CT/Q,[\sigma]}$.
\begin{equation}
\label{orbpartfunc}
Z_{\CT/Q,[\sigma]}=\sum_{\vec g\in Q}\sum_{\vec p\in B_{\vec g}}\chi_{\vec p}(q)\bar\chi_{\overline{\vec p+\vec g}}(q)~,
\end{equation}
where
\be
\label{pconst4}
B_{\vec g}:=\left\{\vec p\ \Big|\ S_{\vec h,\vec  p} ~ \Xi(\vec h,\vec g)=1 ~,\ \forall \vec h \in Q\right\}~.
\ee

A basic observation is that these partition functions are of the form
\begin{eqnarray}
Z_{\CT/Q,[\sigma]}=\sum_{\vec p,\vec q} \CM_{\vec p\vec q}\chi_{\vec p}(q)\bar\chi_{\vec q}(\bar q)~,
\end{eqnarray}
where $\CM_{\vec p\vec q}$ is a modular invariant matrix with entries consisting of 0's and 1's. Indeed, if
\be
\chi_{\vec p}\bar\chi_{\overline{\vec p+\vec g}}=\chi_{\vec q}\bar\chi_{\overline{\vec q+\vec h}}~,
\ee
then we should have $\vec p= \vec q$ and $\vec p+ \vec g= \vec q+ \vec h$ which implies that $\vec g= \vec h$. Therefore, the non-trivial terms contribute to the parition function without mutiplicity.

Now let us discuss some properties of the set $B_{\vec g}$. For any $\vec g$, the set $B_{\vec g}$ is non-empty. To see this, let $K$ be the group defined in equation \eqref{pval}. Let $\{e_i\}$ be a set of generators of this group. Let $\vec h \in Q \lhd K$ be the vector denoting an element of $K$ in the basis $\{e_i\}$. Let $\{f_i\}$ be a basis of $Q$. Then we have
\be
f_i= \sum_j L_{ij} e_j ~,
\ee
for some integer matrix $L$ with non-negative entries. We will focus on $Q=\DZ_2^k$. Therefore, the non-trivial entries of $L_{ij}$ have the form $2^{r_j-1}$. Let $\vec h_Q$ be the vector $\vec h$ written in the basis $\{f_i\}$. Then we have
\be
\vec h= L^T \vec h_Q ~.
\ee
We introduced the basis $\{f_i\}$ because $\Xi$ has a simple description in this basis. It can always be written as
\be
\Xi(\vec h_Q, \vec g_Q)= e^{\pi i \vec h_Q^T X g_Q}~,
\ee
where $X$ is a symmetric integer matrix with diagonal entries equal to $1$ \cite{Fuchs:2004dz}. Now, we have
\be
S_{\vec h,\vec  p} ~ \Xi(\vec h,\vec g)=e^{\pi i \vec h MA \vec p} e^{\pi i \vec h_Q^T X g_Q}= e^{\pi i \vec h_Q^T L M A \vec p}e^{\pi i \vec h_Q^T X g_Q}~.
\ee
Therefore, the constraint \eqref{pconst4} can be simplied to get
\be
 h_Q^T (L M A \vec p+ X \vec g_Q)=0 \text{ mod }2 ~ \forall \vec h_Q \in \DZ_2^k~.
\ee
We get
\be 
L M A \vec p= \vec \alpha - X \vec g_H ~.
\ee
where $\vec \alpha$ satisfies $\vec h_Q \cdot \vec \alpha=0 \text{ mod }2 ~ \forall \vec h_Q \in \DZ_2^k$. This equation always has a solution since $L M A$ is a full rank matrix. Therefore, we find that $B_{\vec g}$ is a non-empty set for all $\vec g$.   

Let us look at how $B_{\vec g}$ are related to $B_{\vec 0}$. For $\vec g=\vec 0$, the constraint \eqref{pconst4} reduces to
\be
S_{\vec h,\vec  p}=1 ~ \forall \vec h \in \DZ_2^k~.
\ee
$B_{\vec 0}$ is the the set of solutions to this constraint. In the bulk TQFT, solutions to this constraint are the Wilson lines which braid trivially with all $\vec h \in \DZ_2^k$. Using Theorem 3.2 in \cite{Muger}, we have
\be
|B_{\vec 0}|= \frac{|K|}{2^k}~.
\ee
Now, given some solution $\vec p\in B_{\vec g}$, $\vec  p+ \vec q \in B_{\vec g}$ where $\vec q \in B_{\vec 0}$. Morevoer, given $\vec p_1, \vec p_2 \in B_{g}$, we have
\bea
S_{\vec h,\vec  p_1} ~ \Xi(\vec h,\vec g)=1= S_{\vec h,\vec  p_2} ~ \Xi(\vec h,\vec g) ~ \forall \vec h \in \DZ_2^k \cr
\implies S_{\vec h, \vec p_1 - \vec p_2}=1~ \forall \vec h \in \DZ_2^k~.
\eea
Therefore, $\vec p_1 - \vec p_2$ belongs to $B_{\vec 0}$. This shows that given some $\vec p \in B_{\vec g}$, all other elements of $B_{\vec g}$ are of the form $\vec p + \vec q$ where $\vec q \in B_{\vec 0}$. Therefore, we have
\be
\label{Bgsize}
|B_{\vec g}|=|B_{\vec 0}|= \frac{|K|}{2^k}~.
\ee

This argument implies that the total number of terms in the partition function $Z_{\CT/Q,[\sigma]}$ is always $|\DZ_2^k| \otimes \frac{|K|}{2^k}=|K|$. Therefore, the map \eqref{codeZ} gives us a code with $2^n$ elements. Hence, the stabilizer code corresponding to the partition function $Z_{\CT/Q,[\sigma]}$ is self-dual. 

\subsection*{Appendix C: Permutation modular invariants and Non-degenerate $\Xi$} \label{AppC}
In this appendix, we prove the following claim:

\noindent
{\bf Claim:} A code CFT is an MPMI if and only if $\Xi(\vec g,\vec h)$, defined in \eqref{SXiZ2k}, is non-degenerate.

\noindent
To understand this claim, let us consider the general expression for the partition function $Z_{\CT/Q,[\sigma]}$.
\begin{equation}
Z_{\CT/Q,[\sigma]}=\sum_{\vec g\in Q}\sum_{\vec p\in B_{\vec g}}\chi_{\vec p}(q)\bar\chi_{\overline{\vec p+\vec g}}(q)~,
\end{equation}
where
\be
\label{pconst3}
B_{\vec g}:=\left\{\vec p\ \Big|\ S_{\vec h,\vec  p} ~ \Xi(\vec h,\vec g)=1 ~,\ \forall \vec h \in Q\right\}~.
\ee
For the partition function to be given by a permutation modular invariant, we know that $\vec p$ as well as $\vec g + \vec p$ should not repeat in the terms of the partition function. Moreover, $\vec p$ should take values in all representations of the chiral algebra. Therefore, it is clear that we need to satisfy the constraint $B_{\vec g} \cap B_{\vec h} = \emptyset$ for $\vec h \neq \vec g \in \DZ_{2}^k$. 

Let us restrict our attention to the case of partition functions which admit a qubit quantum code description. Then we know that the 1-form symmetry $Q=\DZ_{2}^k$ of the bulk TQFT should be anomaly free. Therefore, if $\vec{p} \in B_g$, then $\vec p + \vec g \in B_{\vec g}$. This follows from
\be
S_{\vec h,\vec p + \vec g}= S_{\vec h,\vec  p} S_{\vec h,\vec  g}= S_{\vec h,\vec  p}~,
\ee 
where we have used the fact that $S$ is a bicharacter and $S_{\vec h,\vec  g}=1 ~ \forall \vec h$ since $\DZ_{2}^k$ is anomaly free. Therefore, if $B_{\vec g} \cap B_{\vec h}=\emptyset$ for $\vec h \neq \vec g \in \DZ_{2}^k$, then $\vec g + \vec p$ cannot be the solution to \eqref{pconst3} for some $\vec h \neq g$. Therefore, $\vec g + \vec p$ also does not repeat for different terms in the partition function. This fact, along with \eqref{Bgsize}, then also guarantees that $\vec p$ takes values in all representations. 

Therefore, we find that it is necessary and sufficient to satisfy the constraint 
\be
\label{constper}
B_{\vec g} \cap B_{\vec h}=\emptyset \text{ for } \vec h \neq \vec g \in \DZ_{2}^k
\ee
to have a permutation modular invariant. It is clear from \eqref{pconst3} that if $\Xi(\vec h,\vec g)= \Xi(\vec h,\vec l) ~ \forall \vec h \in \DZ_{2}^k$, then $B_{\vec g}= \vec B_{\vec l}$. Also, suppose $\vec p$ belongs to both $B_{\vec g}$ and $B_{\vec{l}}$. Then using \eqref{pconst3}, we find that $\Xi(\vec h,\vec g)= \Xi(\vec h,\vec l) ~ \forall ~ \vec h \in \DZ_{2}^k$. Therefore, satisfying \eqref{constper} is the same as having a non-degenerate $\Xi(\vec g, \vec h)$.

\subsection*{Appendix D: The Verlinde subgroup} \label{AppD}

Using the results in Appendix C, we know that a non-permutation modular invariant necessarily leads to states of the form $(\vec 0,\bar{\vec g})$ where $\vec g \neq \vec 0$. The states $(\vec 0,\bar{\vec g})$ form a group under fusion we call $E\simeq Z_2^t$. In this appendix, we will discuss how we can extend the chiral algebra using $E$ to get a permutation modular invariant. Then we will discuss how this gives symmetries generated by Verlinde lines which are used to construct the Verlinde subgroup.

To that end, let $\vec \gamma$ denote a representative of the orbit $\{\vec \gamma+ \vec b|\vec b \in E\}$ and $\vec \gamma \in Q$. Now, since $\Xi(\vec h, \vec a)=1$ for any $\vec{a} \in E$ and $\vec{h} \in \DZ_2^k$, $B_{\vec g}= B_{\vec{a}+ \vec{g}}$. That is, $B_{\vec g}$ only depends on the $E$-orbit of $\vec g$. Therefore
\be
Z_{\CT/Q,[\sigma]}=\sum_{\vec \gamma} \sum_{\vec p \in B_{\vec \gamma}} \sum_{\vec b \in E}  \chi_{\vec p}(q) \overline{\chi}_{\overline{\vec p + \vec \gamma+ \vec b}}(\bar q)~,
\ee 
where the subscript on $B_{\vec \gamma}$ indicates that the set of elements in $ B_{\vec g}$ only depends on the $E$-orbit of $\vec g$.

For a given $\vec g$ and $\vec p \in B_{\vec g}$, $\vec p + \vec a$, for any $\vec a \in E$, also belongs to $B_{\vec g}$. This statement follows from that fact that $\vec a,\vec g\in Q$ braid trivially with each other. Therefore, we can put the elements of $B_g$ in orbits under the action of $E$. Let $\vec\rho$ denote the representative of an orbit $\{\vec p + \vec a| \vec a \in E\}$ and $\vec p \in B_{\vec g}$. Then the partition function becomes
\be
Z_{\CT/Q,[\sigma]}=\sum_{\vec\gamma} \sum_{\vec\rho \in B_{\vec \gamma}} \sum_{\vec b \in E} \sum_{\vec a \in E} \chi_{\vec \rho + \vec a} \overline{\chi}_{\overline{\vec \rho + \vec a + \vec G + \vec b}}~.
\ee
In writing this, we have split the sum over $\vec p$ for a given $\vec g$ into a sum over $E$ orbits. We know that $\vec a + \vec b$ is also an element of $E$. Since we are summing over all elements in the group $E$, we can change variables and obtain
\bea
Z_{\CT/Q,[\sigma]}&=&\sum_{\vec \gamma} \sum_{\vec \rho \in B_{\vec \gamma}} \sum_{\vec b \in E} \sum_{\vec a \in E} \chi_{\vec \rho + \vec a} \overline{\chi}_{\overline{\vec \rho + \vec \gamma + \vec b}}=\cr &=&\sum_{\vec \gamma} \sum_{\vec \rho \in B_{\vec \gamma}}  \bigg (\sum_{\vec a \in E} \chi_{\vec \rho + \vec a}\bigg ) \bigg ( \sum_{\vec b \in E} \overline{\chi}_{\overline{\vec \rho  + \vec \gamma + \vec b}} \bigg )~.\ \ \ \ \ \ 
\eea

Therefore, we can enlarge the chiral algebra where the vaccum character of the new chiral algebra is given by $\tilde \chi_{\vec 0}:= \sum_{\vec a \in E} \chi_{\vec a}$ and, more generally 
\be
\label{charactermap}
\tilde \chi_{\vec \rho}:= \sum_{\vec a \in E} \chi_{\vec \rho + \vec a}~.
\ee
Then the partition function becomes
\be\label{permLC}
Z_{\CT/Q,[\sigma]}=\sum_{\vec \gamma} \sum_{\vec \rho \in B_{\vec \gamma}} \tilde \chi_{\vec{\rho}} \overline{\tilde \chi}_{\overline{\vec{\rho} + \vec \gamma}}~.
\ee

In fact, this is again a permutation modular invariant. To see this, let $\vec \delta$ and $\vec \epsilon$ lie in two distinct $E$-orbits. Then the sets $B_{\vec \delta}$ and $B_{\vec \epsilon}$ have no common elements since otherwise (using the bi-character nature of $\Xi$)
\be
\Xi(\vec h, \vec \delta)= \Xi(\vec h, \vec \epsilon) ~, \forall \vec h \in \DZ_2^k\ \Rightarrow\ \Xi(\vec h, \vec \delta + \vec \epsilon)=1~.
\ee
Therefore, $\vec \delta + \vec \epsilon$ would be an element of $E$ which would imply that $\vec \delta$ and $\vec \epsilon$ are in the same $E$-orbit (a contradiction). Therefore, for every $\vec \gamma$, the sum over $\vec\rho$ is over elements which do not repeat for any $\vec \eta \neq \vec \gamma$. Also, we know that $\vec \rho + \vec \gamma \in B_{\vec \gamma}$ if $\vec \rho \in B_{\vec \gamma}$. As a result, in \eqref{permLC}, the values of $\overline{\vec \rho + \vec \gamma}$ do no repeat either. In other words, after enlarging the chiral algebra, we end up with a permutation modular invariant theory with respect to this new chiral algebra.

It is now clear that we have Verlinde lines labelled by primaries with respect to the enlarged chiral algebra. Then, consider the following defect partition function 
\be
Z_{\CT/Q,[\sigma]}^{\vec \zeta}= \sum_{\vec \gamma} \sum_{\vec \rho \in B_{\vec \gamma}} \tilde \chi_{\vec{\rho} + \vec \zeta} \overline{\tilde \chi}_{\overline{\vec{\rho} + \vec \gamma}}~.
\ee
To get a map to the corresponding code elements, it is easier to use \eqref{charactermap} and substitute
\bea
Z^{\vec \zeta}_{\CT/Q,[\sigma]}&=&\sum_{\vec \gamma} \sum_{\vec \rho \in B_{\vec \gamma}} \bigg (\sum_{\vec a \in E} \chi_{\vec \rho + \vec \zeta + \vec a} \bigg ) \bigg (\sum_{\vec b \in E} \chi_{\vec \rho + \vec \gamma + \vec b} \bigg )\cr&=&\sum_{\vec \gamma} \sum_{\vec \rho \in B_{\vec \gamma}} \sum_{\vec a, \vec b \in E}  \chi_{(\vec \rho + \vec a) + \vec \zeta} \chi_{(\vec \rho + \vec a) + \vec \gamma + \vec a + \vec b}~.
\eea
When we sum over $\vec a \in E$, the term $\vec \rho + \vec a$ runs over the E-orbit of $\vec\rho \in B_{\vec\gamma}$. Also, the term $\vec a + \vec b$ is just a permutation of $\vec b$. Since we are summing over all $\vec b \in E$ as well, we can simplify the expression above to get
\be\label{verF}
Z^{\vec \zeta}_{\CT/Q,[\sigma]}=\sum_{\vec \gamma} \sum_{\vec p \in B_{\vec \gamma}} \sum_{\vec b \in E}  \chi_{\vec p + \vec \zeta} \chi_{\vec p + \vec \gamma + \vec b} = \sum_{\vec g} \sum_{\vec p \in B_{\vec g}}  \chi_{\vec p + \vec \zeta} \chi_{\vec p + \vec g}~.
\ee
Note that $\vec \zeta$ need not be an order-two element of the MTC of the original chiral algebra, even though it may be an order-two element in the MTC of the extended chiral algebra. In fact, if $\vec \zeta$ is not an order-two element of the original MTC, then we cannot relate the defect operators $\{\CO^{\vec \zeta}_{\vec p+ \vec \zeta,\vec p + \vec g}\}$ to a Pauli group element. If $\vec \zeta$ is order two, then from the terms in \eqref{verF}, we get the Pauli group elements
\be
\{\CO^{\vec \zeta}_{\vec p+ \vec \zeta,\vec p + \vec g}\} \leftrightarrow X^{M(\vec g + \vec \zeta)} \circ Z^{A(\vec p + \vec \zeta)}~.
\ee
Note that here $\vec p \in B_{\vec g}$ is not independent of $\vec g$. In general, our RCFTs will have other sources of order-two lines that furnish the remainder of the Pauli group (as discussed in Section \ref{defects}). In the extreme example of theories that are modular-invariant holomorphic RCFTs times modular-invariant anti-holomorphic RCFTs, all order-two lines are non-Verlinde lines.

Since the Verlinde subgroup $\mathcal{V}_{\CT/Q}$ is formed by order two elements, it is isomorphic to $\DZ_2^{N_v}$. Here $N_v$ is the number of Pauli group elements obtained from the defect partition functions \eqref{verF}. In general $|\mathcal{V}_{\CT/Q}|$ will depend on the choice of the group $Q$ by which we orbifold the CFT with the charge-conjugation partition function to get $Z_{\CT/Q,[\sigma]}$. But when the group $K$ defined in \eqref{pval} is such that $n_{A_2}=n_{B_2}=n_{C_2}=n_{D_2}=n_{E_2}=n_{F_2}=0$, then we can find a general expression for $|\mathcal{V}_{\CT/Q}|$. This constraint is the same as imposing that $K$ does not have any $\DZ_2$ factors. Note that we also ignore decoupled CFT factors described by $A_{q^r}$ and $B_{q^r}$. 

Consider the general expression of the $S$ matrix $S_{\vec p, \vec q}=e^{\frac{2 \pi i}{2} \vec p^T M A \vec q}$. Consider an element $\vec p \in B_{\vec 0}$ which satisfies
\be
\label{constappD}
S_{\vec h, \vec p}=1 \forall \vec h \in Q=\DZ_2^k \implies \vec h^T M A \vec p=0 \text{ mod }2~.
\ee
Note that since $\vec h$ is an order two vector, $h^T M$ is an integer vector. Moreover, $\vec h$ has even components. $A$ is also an integer matrix by definition. Let $\vec p$ be an order two vector. Then, it has even components. This follows from our assumption that $K$ does not have any $\DZ_2$ factors. Therefore, any order two vector satisfies the constraint \eqref{constappD}. That is, all the $2^n$ distinct order two elements belong to $B_{\vec 0}$, where $n$ is the number of qubits in the corresponding quantum code or equivalently the length of the vector $\vec p$. 

When we enlarge the chiral algebra to obtain a permutaion modualar invariant, these $2^n$ order two elements are put into orbits under the group $E$. Each such orbit defines a Verlinde line whose defect partition function gives $2^n$ Pauli group elements. This follows form the fact that the partition function itself gives $2^n$ distinct stabilizer elements, as we showed in Appendix B.  Therefore, the size of the Verlinde subgroup is
\be
2^{n-t} \times 2^{n}~,
\ee
where $|E|=2^t$. If the Schellekens algebra gives a permutation modular invariant, $t=0$ and the Verlinde subgroup has size $2^n \times 2^n=4^n$. Therefore, we get the full Pauli group.

\end{document}